\documentclass[12pt]{article}
\usepackage{pdfpages}
\renewcommand{\theequation}{\arabic{section}.\arabic{equation}}
 
\def\be{\begin{equation}}
\def\ee{\end{equation}}
\def\bqa{\begin{eqnarray}}
\def\eqa{\end{eqnarray}}
\def\lsim{\raise0.3ex\hbox{$<$\kern-0.75em\raise-1.1ex\hbox{$\sim$}}}
\def\gsim{\raise0.3ex\hbox{$>$\kern-0.75em\raise-1.1ex\hbox{$\sim$}}}
\usepackage{amssymb}

\begin{document}

\begin{center}
{\bf Color Transparency and Saturation in QCD\footnote{Presented at the 
International School of Subnuclear Physics, 50th Course, Erice, Sicily,
23 June -- 2 July 2012}}
\end{center}
\vspace {0.5cm}
\begin{center}
{\bf Dieter Schildknecht\footnote{Email: schild@physik.uni-bielefeld.de}}
\\[2.5mm]
Fakult\"at f\"ur Physik, Universit\"at Bielefeld,\\[1.2mm]
  Universit\"atsstra\ss e 25, 33615 Bielefeld, Germany\\[1.2mm]
and\\[1.2mm]
Max-Planck-Institute for Physics, F\"ohringer Ring 6, \\[1.2mm]
80805 Munich, Germany
\end{center} 


\begin{center}
{\bf Abstract}
\end{center}

We review the theoretical interpretation of deep-inelastic electron-proton
scattering at low values of the Bjorken variable $x \simeq Q^2/W^2 \lsim 0.1$.
The process proceeds via the interaction of quark-antiquark
$(q \bar q)$ color-dipole
fluctuations of the (virtual) photon with the proton. In terms of the forward
Compton scattering amplitude, two reaction channels contribute to the
interaction of the $q \bar q$ color dipole with the gluon field in the
proton. Dependent on the kinematics, there is either {\it color transparency},
corresponding to a cancellation of the amplitudes for the two reaction
channels, or {\it saturation}, occuring when the process is dominated by 
a single interaction channel. The connection between the color-dipole
picture and the pQCD improved parton model is elaborated upon.

\section{Introduction}
\label{}

The concepts of {\it color transparency} and {\it saturation} in QCD
became relevant in connection with the interpretation
of the deep inelastic scattering (DIS) experiments carried out at
the $e^\pm$-proton collider in Hamburg in the years from 1992 to
2007. The available center-of-mass energy $W$ allowed to explore
DIS in a previously non-accessible range of very low values of the
Bjorken variable $x \equiv x_{bj} \simeq Q^2/W^2$,
\be
5 \cdot 10^{-4} \le x \le 10^{-1},
\label{1.1}
\ee
combined with a nevertheless appreciable range of the four-momentum 
squared of the virtual photon, approximately given by
\be
0 \le Q^2 \le 100 GeV^2.
\label{1.2}
\ee

One of the outstanding findings from the HERA experiments is the
low-x scaling behavior of the photoabsorption cross section,
$\sigma_{\gamma^*p} (W^2,Q^2)$. When plotting, compare Fig.\,1, the 
experimental
data for $\sigma_{\gamma^*p} (W^2,Q^2)$ as a function of the low-x
scaling variable
\be
\eta (W^2,Q^2) = \frac{Q^2 + m^2_0}{\Lambda^2_{sat} (W^2)},
\label{1.3}
\ee
one finds a single curve \cite{DIFF2000, SCHI},
\be
\sigma_{\gamma^*p} (W^2,Q^2) = \sigma_{\gamma^*p} (\eta(W^2,Q^2)) \sim
\sigma^{(\infty)}
\left\{ \matrix{
\frac{1}{\eta (W^2, Q^2)},~~~~~~{\rm for}~~(\eta (W^2, Q^2) \gg 1), \cr
\ln \frac{1}{\eta (W^2,Q^2)},
~~~{\rm for}~~ (\eta (W^2,Q^2) \ll 1). } \right.
\label{1.4}
\ee
In (\ref{1.3}), the ``saturation scale'' $\Lambda^2_{sat} (W^2)$ rises with
a small constant power, $C_2$, of the energy, 
\be
\Lambda^2_{sat} (W^2) \sim (W^2)^{C_2}.
\label{1.5}
\ee
\begin{figure}[h]
\begin{center}
\includegraphics[scale=.35]{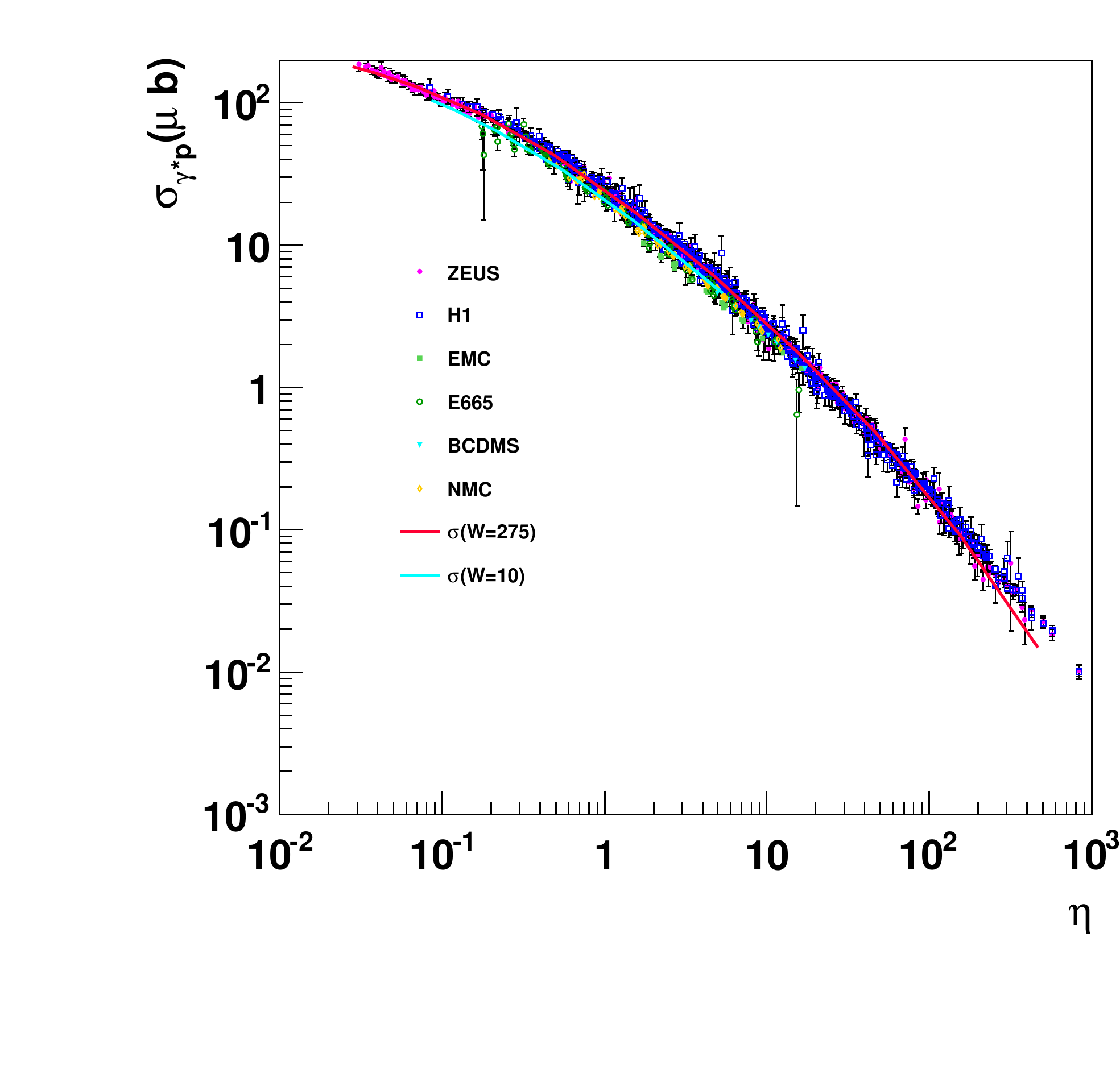}
\vspace*{-1.2cm}
\caption{Scaling \cite{DIFF2000, SCHI} of $\sigma_{\gamma^*p} 
(W^2,Q^2) = \sigma_{\gamma^*p} (\eta
(W^2,Q^2))$.}
\vspace*{-0.5cm}
\end{center}
\end{figure}
The value of $m_0$ is somewhat smaller than the mass of the $\rho^0$
meson. The quantity $\sigma^{(\infty)}$ denotes a cross section that in
good approximation is independent of the photon energy. The log-log plot
of the experimental data in Fig. 1 immediately allows one to infer the
$1/\eta (W^2,Q^2)$ dependence for $\eta (W^2,Q^2)\gg 1$ in (\ref{1.4}),
as well as the $\ln (1/\eta(W^2,Q^2)$ dependence on $\eta (W^2,Q^2)$ for
$\eta (W^2,Q^2) \ll 1$.

The logarithmic behavior in the limit of $\eta(W^2,Q^2) \to 0$, or
$W^2 \to \infty$ at $Q^2$ fixed, leads to the conclusion that the
ratio of the virtual to the real photoabsorption cross section tends
towards unity in this ``saturation'' limit \cite{DIFF2000, SCHI},
\be
\lim_{W^2 \to \infty \atop {Q^2~{\rm fixed}}} 
\frac{\sigma_{\gamma^*p} (\eta (W^2, Q^2))}{\sigma_{\gamma^*p} (\eta
(W^2, Q^2 = 0))} 
= \lim_{W^2 \to \infty \atop {Q^2~{\rm fixed}}} 
\frac{\ln \left( \frac{\Lambda^2_{sat}(W^2)}{m^2_0} 
\frac{m^2_0}{(Q^2 + m^2_0)} \right)}{\ln 
\frac{\Lambda^2_{sat} (W^2)}{m^2_0}} 
= 1 + \lim_{W^2 \to \infty \atop {Q^2~{\rm fixed}}} 
\frac{\ln \frac{m^2_0}{Q^2 + m^2_0}}{\ln 
\frac{\Lambda^2_{sat} (W^2)}{m^2_0}} = 1.
\label{1.6}
\ee

The dependence of the photoabsorption cross section in 
(\ref{1.4}) and (\ref{1.6}) corresponds to a simple subdivision of the 
$(Q^2,W^2)$ plane. It is shown in Fig.\,2.
\begin{figure}[h]
\begin{center}
\vspace*{-0.5cm}
\includegraphics[scale=.35]{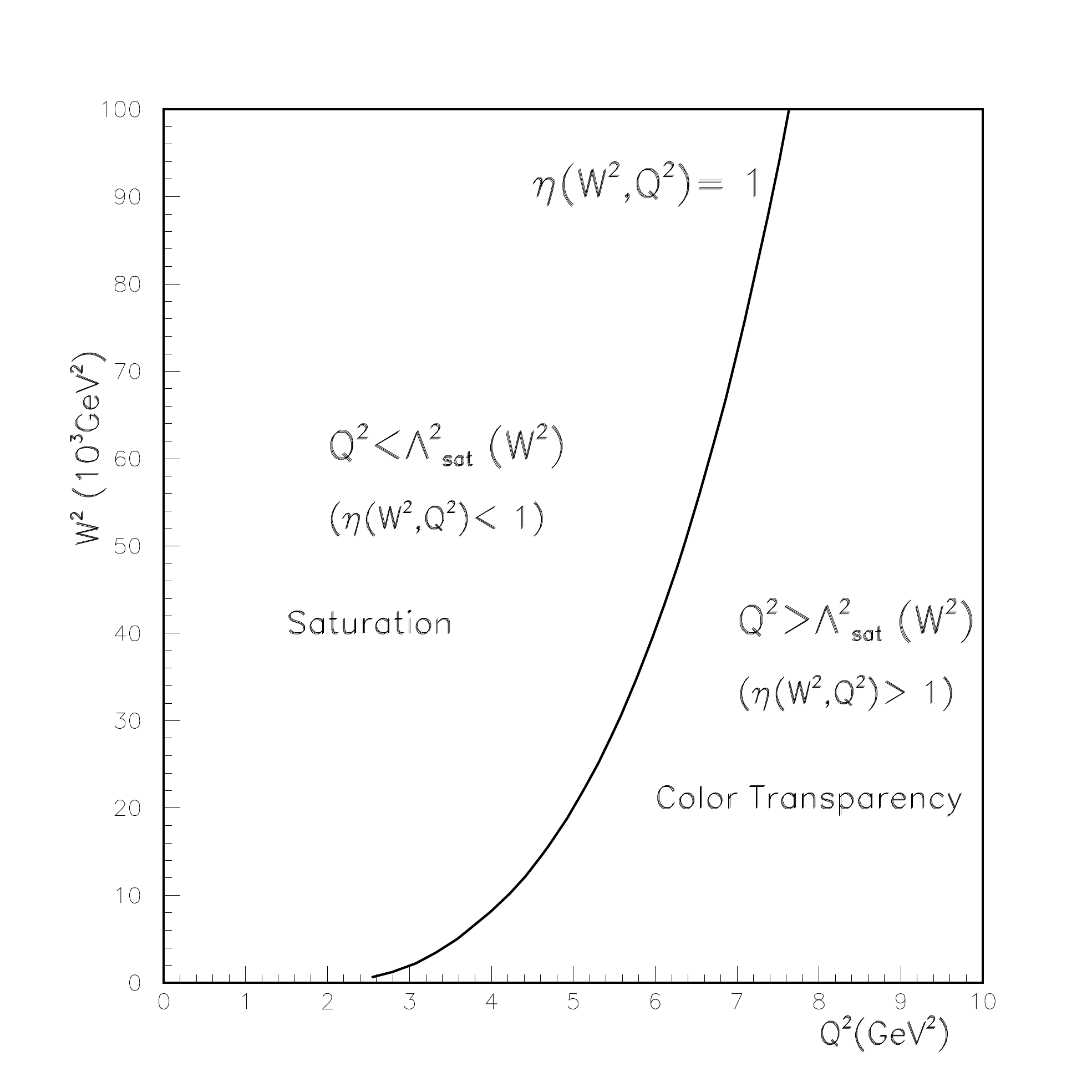}
\caption{The $(Q^2,W^2)$ plane corresponding to the scaling behavior
of $\sigma_{\gamma^*p} (W^2,Q^2) = \sigma_{\gamma^*p} (\eta (W^2,Q^2))$
shown in Fig.\,1. The line $\eta (W^2, Q^2) = 1$
subdivides the $(Q^2,W^2)$ plane into the saturation region of $\eta (W^2,Q^2)
< 1$ and the color transparency region of $\eta (W^2,Q^2) > 1$.}
\end{center}
\vspace*{-0.8cm}
\end{figure}
The region of $\eta (W^2,Q^2) \ll 1$, corresponds to ``saturation'' 
in the sense of an
approach to the $Q^2 = 0$ photoabsorption cross section, while the region of
$\eta (W^2,Q^2) \gg 1$ corresponds to ``color transparency''. The choice
of this latter terminology will be justified in Section 3 below.

It is the aim of the present lecture to show that the scaling behavior with
the functional form in (\ref{1.4}) arises as a consequence of the
color-gauge-invariant interaction of $q \bar q$ color-dipole fluctuations
of the photon with the proton (color-dipole picture, (CDP)). No specific
parameter-dependent assumption on the dipole-proton interaction is
needed to arrive at this conclusion. 

Again, without parameter adjustment, the CDP leads to a unique prediction
\cite{Ku-Schi, E}
for the ratio $R(W^2,Q^2)$ of the longitudinal to the transverse 
photoabsorption cross section at sufficiently large $Q^2 > \Lambda^2_{sat}
(W^2)$. The definite value of $R(W^2,Q^2)$ arises as a consequence of 
the different transition probabilities for longitudinally polarized and
transversely polarized photons into $q \bar q$ states, combined with the
different transverse interaction size of $q \bar q$ states originating
from longitudinally versus transversely polarized photons.

Consistency of the CDP with the pQCD-improved parton model allows one to
deduce \cite{SCHI, N9, 23a,E}
the numerical value of the exponent $C_2$ of the saturation scale
$\Lambda^2_{sat} (W^2)$. 

It is sometimes argued that the DIS data provide
empirical evidence for parton recombination requiring non-linear parton
evolution mechanisms (compare e.g. ref. \cite{F} and the list of references
therein). We do not see such an evidence.

\section{Photon hadron interactions: Late 1960's and early 1970's}
\renewcommand{\theequation}{\arabic{section}.\arabic{equation}}
\setcounter{equation}{0}

The general features of our present understanding of photon-hadron
interactions at low values of $x \simeq Q^2/W^2 \ll 0.1$ date back to
the pre-QCD era of the late 1960's and early 1970's. So let me briefly
return to that time.

Photon-hadron interactions in the late 1960's and early 1970's were
dominated by the vector-meson-dominance (VMD) picture \cite{N1}-\cite{A}: 
the photon
virtually dissociates, or fluctuates in modern jargon, into the
low-lying vector mesons, $\rho^0, \omega, \phi$, that subsequently
interact with e.g. the proton. Compare Fig.\,3, showing the specific
case of (diffractive) vector meson production in photon-proton
interactions. Inserting the $\gamma (\rho^0, \omega, \phi)$
\begin{figure}[h]
\begin{center}
\includegraphics[scale=.6]{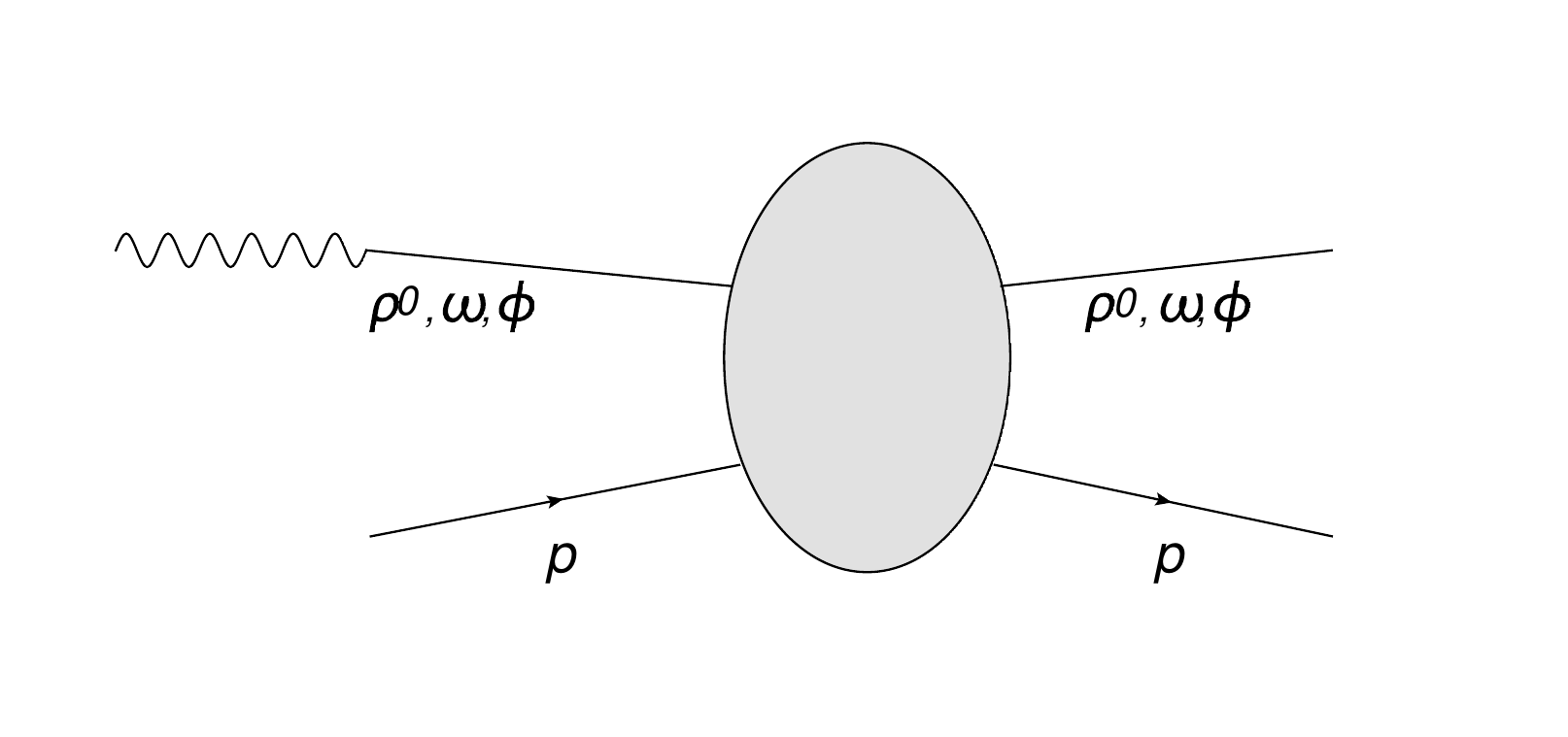}
\vspace*{-0.8cm}
\caption{Vector-meson production in photon-proton interactions.}
\end{center}
\vspace*{-0.7cm}
\end{figure}
couplings that had become available from $e^+e^-$-annihilation experiments
in the late 1960's, and relating the vector-meson scattering amplitudes to
the ones from pion-nucleon scattering by invoking the additive quark model,
vector meson production was quantitatively predicted as a function of
energy and momentum transfer. The total photoproduction cross section
then followed from the optical theorem by attaching a photon to the
final vector mesons in Fig.\,3. The experimental findings were theoretically
summarized in terms of a ``hadronlike behavior'' 
\cite{N3, N4} of the photon.

Based on vector-meson dominance, hadronlike behavior of the cross section
of photoproduction on heavy nuclei was predicted \cite{B}
as a consequence of the
destructive interference of the one- and two-channel interactions depicted
in Fig.\,4. Experiments at DESY and SLAC confirmed the theoretical prediction
of a hadronlike dependence on the mass number of the nucleus (``shadowing'').
\begin{figure}[h]
\begin{center}
\vspace*{-0.5cm}
\includegraphics[scale=.6]{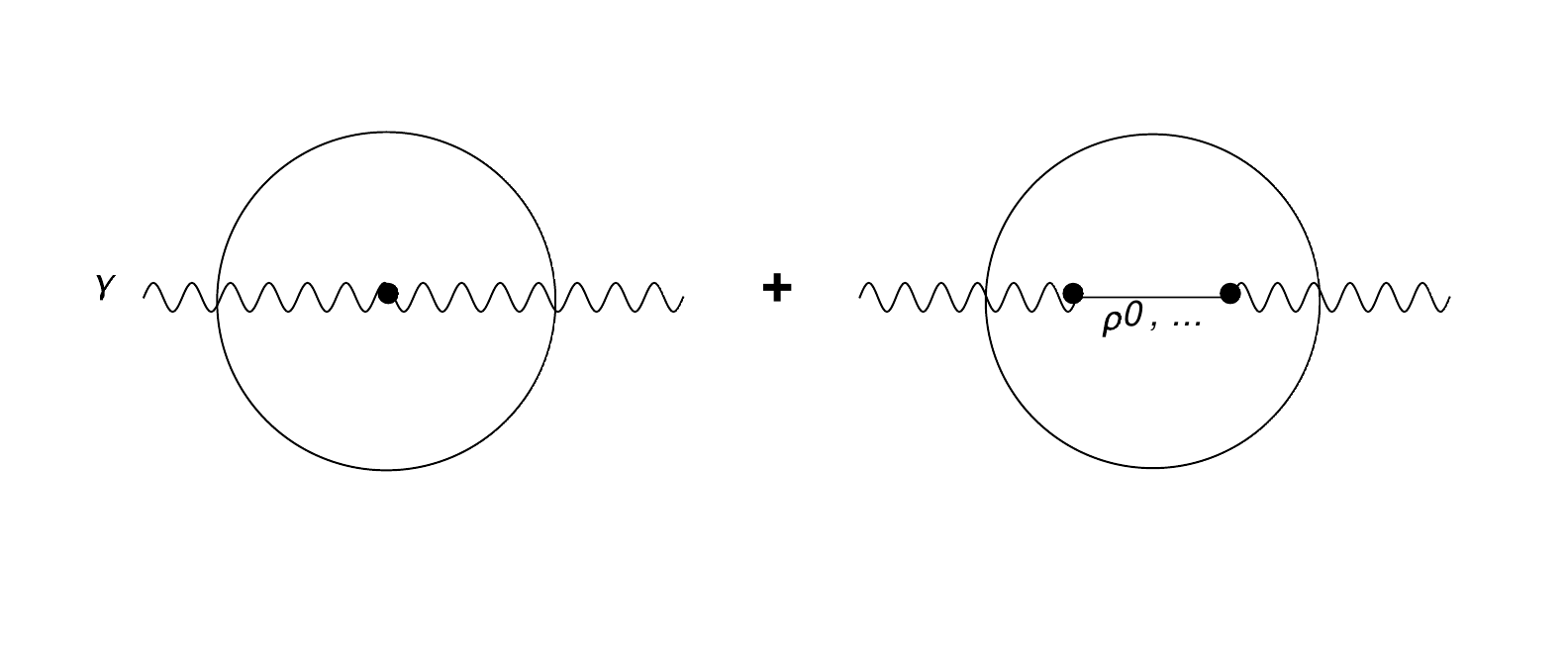}
\vspace*{-1cm}
\caption{The forward-Compton-scattering amplitude of the interaction
of a photon with a heavy nucleus.}
\end{center}
\vspace*{-0.5cm}
\end{figure}

The picture of photon-hadron interactions changed dramatically, when,
starting in 1969, the first experimental results on DIS at large values
of $Q^2$ from the SLAC-MIT collaboration became available \cite{N5}. The
experimental results provided evidence for the scaling behavior of the 
proton structure
function that had been
conjectured by Bjorken \cite{N6}, and they led Feynman to propose
the parton-model \cite{N7}.

An alternative point of view concerning the results from the SLAC-MIT
experiment was advocated by Sakurai and myself in 1972 \cite{Sakurai}. 
Since this point
of view is of much relevance as a predecessor of our present understanding
of DIS at low $x \equiv x_{bj}$ in QCD, I briefly remind you of our 
1972 procedure.

We conjectured that the slow decrease of the photoabsorption cross section
with increasing $Q^2$ observed by the SLAC-MIT collaboration at large
$w^\prime \sim 1/x$ was due to the, at that time 
hypothetical\footnote{First experimental data on $e^+e^-$ annihilation
beyond the $\rho^0, \omega, \phi$ region had become available 
\cite{Bernardini} from
the Frascati $e^+e^-$ storage ring (Adone) in 1971, providing evidence
for production of multi-hadron final states at 1.3 GeV to
1.4 GeV $e^+ e^-$ center-of-mass energy.}, coupling
of the photon to a high-mass continuum predicted to be observed in future
$e^+e^-$ annihilation experiments beyond the $\rho^0, \omega, \phi$ region.
Compare Fig.\,5
\begin{figure}[h]
\begin{center}
\includegraphics[scale=.6]{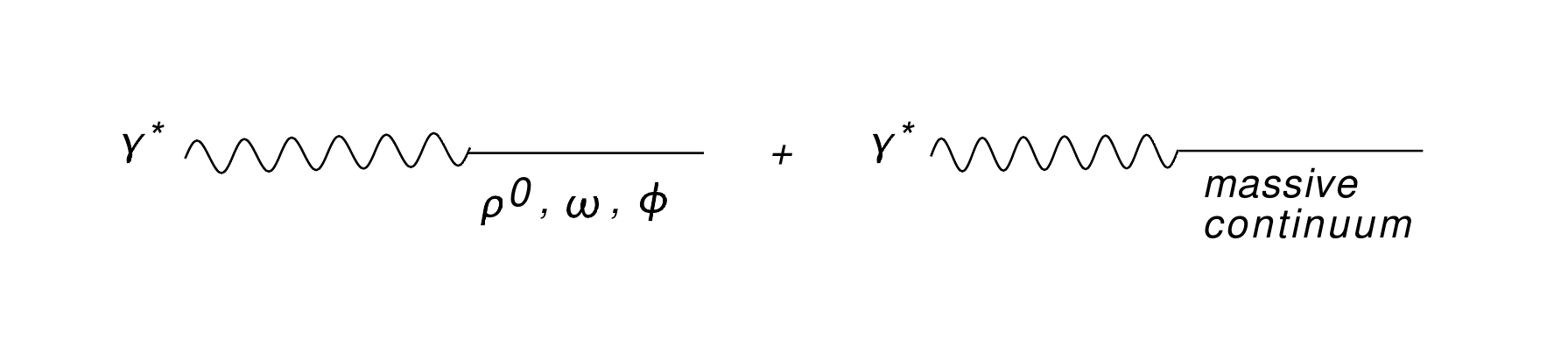}
\vspace*{-0.5cm}
\caption{The coupling of the photon to a high-mass continuum.}
\end{center}
\end{figure}
\begin{figure}[h]
\begin{center}
\includegraphics[scale=.6]{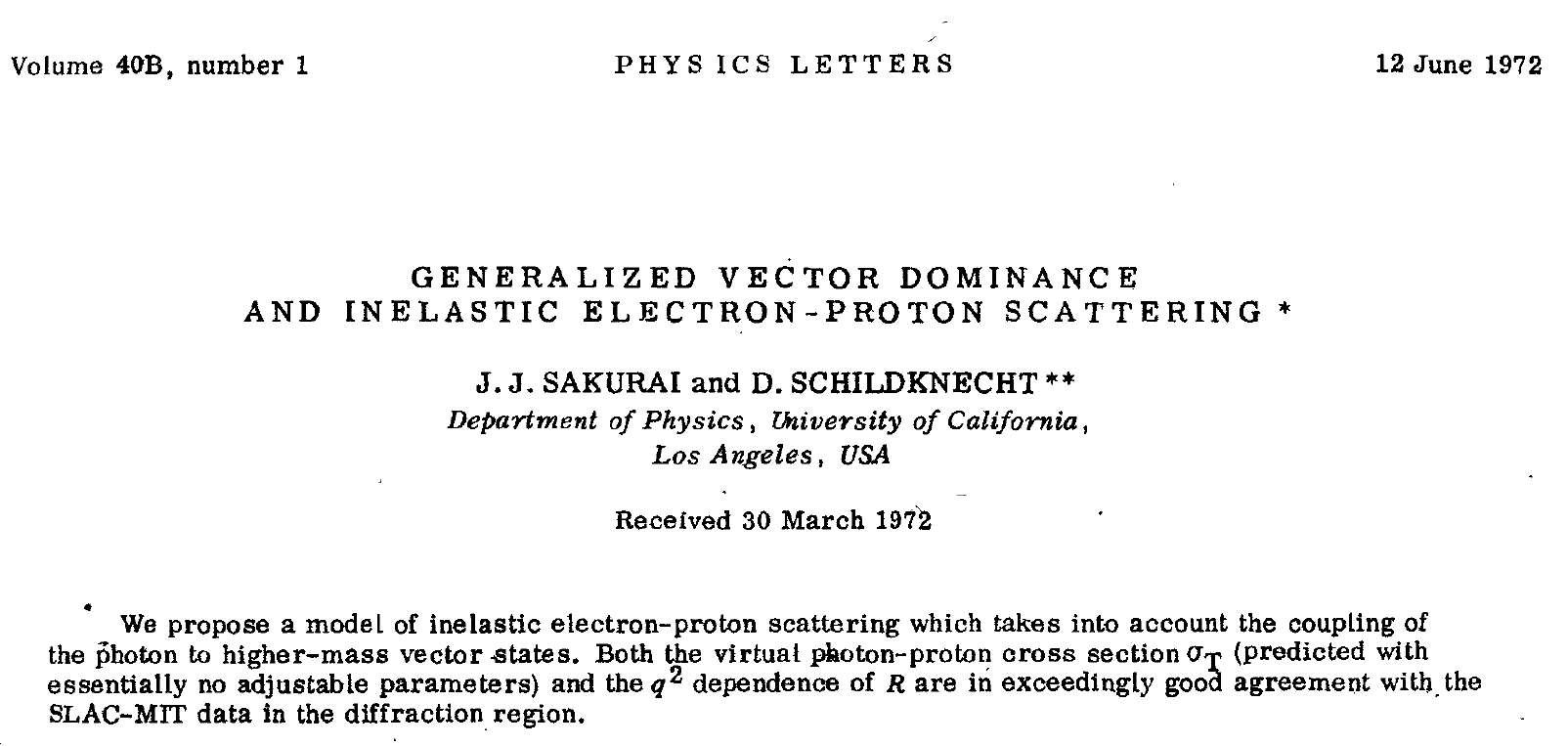}
\end{center}
\end{figure}
and the comparison of our generalized vector dominance (GVD) predictions
\cite{Sakurai} with the experimental data in Fig.\,6. The GVD approach was
further developed in ref. \cite{Fraas}, compare also \cite{N10} and
the review in ref. \cite{A}.
\begin{figure}[h!]
\begin{center}
\includegraphics[scale=.6]{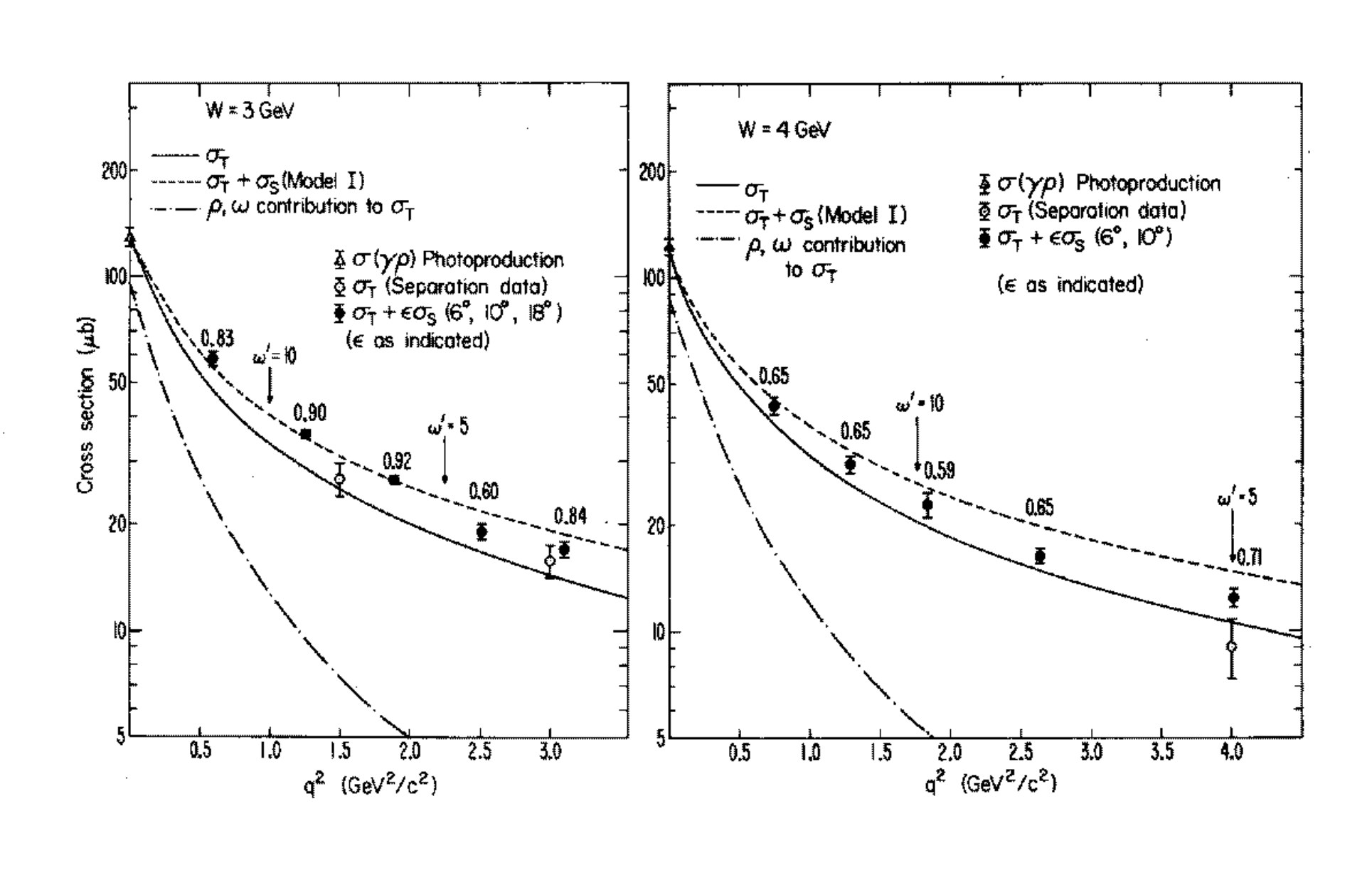}
\vspace*{-0.5cm}
\caption{The SLAC-MIT experimental data compared with GVD predictions 
\cite{Sakurai}.}
\end{center}
\end{figure}

In the early 1970's, it was frequently argued that our GVD interpretation of
the experimental results was incorrect, since experiments did not reveal
the hadronlike shadowing in the scattering of virtual photons from complex
nuclei implied by GVD. After many years of confusion, this point was clarified
by the 1989 experimental results of the EMC-NMC collaboration \cite{C}, 
compare Fig.\,7.
The results indeed showed the shadowing effect expected for virtual
photons due to high-mass fluctuations \cite{D}
of the virtual photon that contribute
to the two-step interaction in Fig.\,4.

\begin{figure}[h]
\begin{center}
\includegraphics[scale=.4]{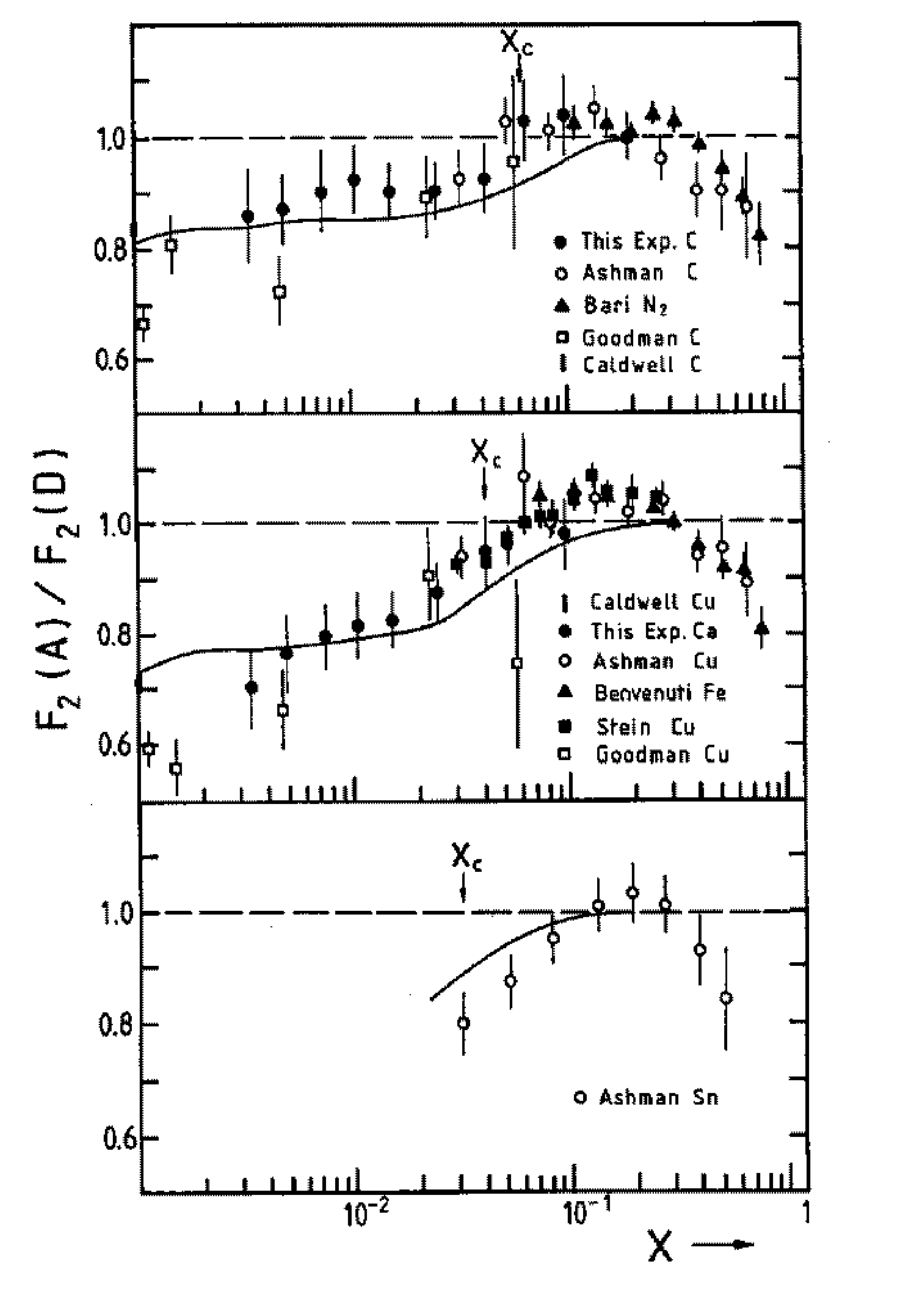}
\caption{Shadowing in the scattering of virtual photons from complex nuclei
compared with GVD predictions \cite{D}}
\end{center}
\end{figure}

Since shadowing for virtual photons requires high-mass diffractive production,
contrary to what is often stated, it came without surprise that one of the
first results from HERA was the observation of ``rap-gap'' events 
\cite{N11} due to
diffractive production of high-mass hadron states.

The 1972 approach of GVD \cite{Sakurai} was based on postulating 
a mass dispersion
relation as immediate generalization of $(\rho^0, \omega, \phi)$-dominance,
\be
\sigma_{\gamma^*_Tp} (W^2,Q^2) = \int dm^2 \int dm^{\prime 2}
\frac{\rho (W^2,m^2,m^{\prime 2}) m^2 m^{\prime 2}}
{(Q^2+m^2)(Q^2+m^{\prime 2})}.
\label{2new.1}
\ee
The momentum space dispersion-theoretic ansatz may be equivalently
interpreted in terms of a three-momentum conserving 
energy-conservation-violating transition to an on-shell hadronic state 
of finite, but sufficiently long lifetime given by \cite{N12, D}
\be
\tau = \frac{1}{\Delta E} = \frac{2 M\nu}{Q^2 + M^2_{q \bar q}}
\frac{1}{M_p} = \frac{1}{x + \frac{M^2_{q \bar q}}{W^2}} \frac{1}{M_p}
\gg \frac{1}{M_p},
\label{2new.2}
\ee
followed by the interaction of the hadronic fluctuation with the nucleon.
In (\ref{2new.2}), $\nu$ denotes the proton-rest-frame energy of the photon and
$M_{q \bar q}$ and $M_p$ denote the fluctuation mass and the proton mass,
respectively.

\section{The modern picture of DIS at low x: the color dipole picture}
\renewcommand{\theequation}{\arabic{section}.\arabic{equation}}
\setcounter{equation}{0}

As in GVD, in the modern approach, the photoabsorption
reaction proceeds in two steps, i) the
$\gamma^* \to q \bar q$ transition and ii) the $(q \bar q)$-proton interaction,
whereby taking into account
\begin{itemize}
\item[i)] the internal structure of the $q \bar q$ system via the variable $0
  \le z \le 1$ that specifies the longitudinal momentum distribution between
  the quark and the antiquark the photon fluctuates into, compare Fig.\,8, and
\begin{figure}[h]
\begin{center}
\vspace*{-0.3cm}
\includegraphics[scale=.6]{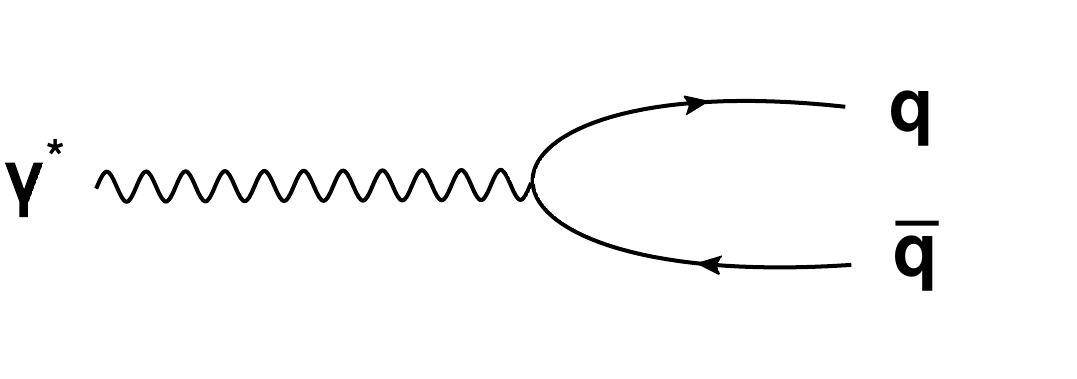}
\vspace*{-0.5cm}
\caption{The $\gamma^* q \bar q$ transition}
\end{center}
\vspace*{-1cm}
\end{figure}
\item[ii)] the $q \bar q$ interaction with the gluon field in the nucleon 
 \cite{Low}  as a
  gauge-invariant color-dipole interaction, compare the
  (virtual) forward Compton scattering amplitude in Fig.\,9.
\end{itemize}
\begin{figure}[h]
\begin{center}
\includegraphics[scale=.8]{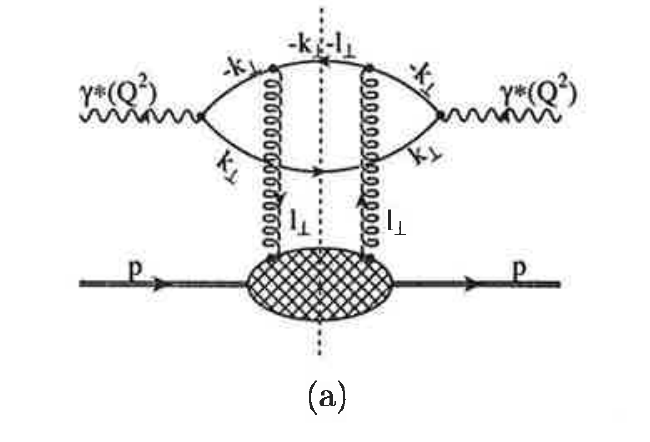}\hspace*{1cm}
\includegraphics[scale=.8]{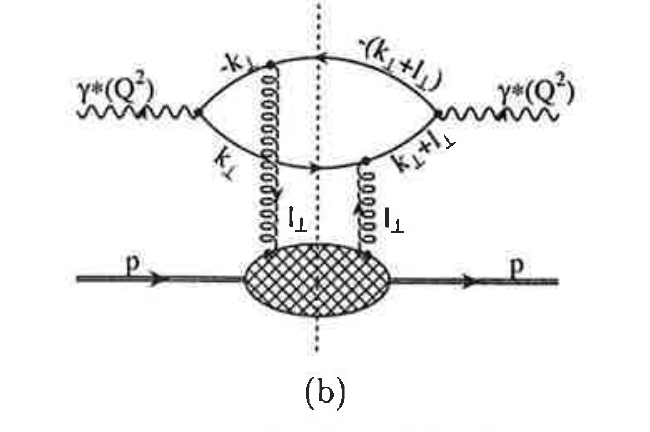}
\vspace*{-0.3cm}
\caption{Two of the four diagrams for the $q \bar q$ dipole interaction with
the gluon field in the nucleon. The diagrams (a) and (b) correspond to
channel 1 and channel 2 respectively.}
\end{center}
\vspace*{-0.6cm}
\end{figure}

After Fourier transform to transverse position space, the photoabsorption 
cross section at low $x \simeq Q^2/W^2 < 0.1$ takes the form \cite{Nikolaev,
Cvetic}
\be
\sigma_{\gamma^*_{L,T}} (W^2, Q^2)  = \int dz \int d^2 \vec r_\bot
\vert \psi_{L,T} (\vec r_\bot, z (1 - z), Q^2) \vert^2 \cdot
\sigma_{(q\bar q)p}
(\vec r_\bot, z (1 - z), W^2). 
\label{2.1}
\ee
The quantity $\vert \psi_{L,T} (\vec r_\bot, z (1-z),Q^2) \vert^2$ may be
interpreted as the probability for a longitudinally or a transversely polarized
photon, $\gamma^*_{L,T}$, of virtuality $Q^2$ to undergo a transition to a
$q \bar q$ state, $\gamma^*_{L,T} \to q \bar q$, being characterized by the
transverse size $\vec r_\bot$ and by the distribution of the longitudinal
momenta of quark and antiquark determined by $z(1-z)$. In the rest frame of
the $q \bar q$ fluctuation of mass $M_{q \bar q}$, the quantity $z(1-z)$ 
determines \cite{Cvetic} the direction of the quark (antiquark) with respect 
to the photon direction. The interaction cross section of the 
$q \bar q$ dipole state in
(\ref{2.1}) 
is denoted by $\sigma_{(q \bar q)p} (\vec r_\bot, z(1-z), W^2)$. It 
depends on the energy, $W$, \cite{Cvetic, DIFF2000, Forshaw, Ewerz} of the
$(q \bar q)p$ interaction, since the photon fluctuates into an on-shell
$q \bar q$ state of mass $M_{q \bar q}$ that subsequently interacts with
the nucleon. For generality, a dependence on $z (1-z)$ is allowed for in
the dipole-proton cross section.

The gauge-invariant two-gluon interaction ii) of the $q \bar q$ dipole enters
the photoabsorption cross section in (\ref{2.1}) via \cite{Nikolaev, Cvetic}
\be
\sigma_{(q \bar q)p} (\vec r_\bot, z(1-z), W^2) = 
\int d^2 \vec l_\bot \tilde{\sigma}
(\vec l^{~2}_\bot, z(1-z), W^2) 
\left(1-e^{-i~ \vec l_\bot \cdot \vec r_\bot}\right).
\label{2.2}
\ee
In (\ref{2.2}), $\vec l_\bot$ stands for the transverse momentum of the
absorbed gluon, and the first and the second term in the parenthesis on the
right-hand side in (\ref{2.2}), respectively, corresponds to the first and the
second diagram in Fig.\,9.

For the subsequent discussions, it will be useful to equivalently 
rewrite \cite{MKDS, E} the
photoabsorption cross section (\ref{2.1}), in terms of dipole states, 
$(q \bar q)^{J=1}_{L,T}$, that describe longitudinally and 
transversely polarized $q \bar q$ states of fixed spin $J=1$ and polarization
index $L$ and $T$. In terms of the corresponding dipole cross section, 
$\sigma_{(q \bar q)^{J=1}_{L,T}p} (\vec r^{~\prime}_\bot, W^2)$, where
$\vec r^{~\prime}_\bot = \vec r_\bot z (1-z)$, the photoabsorption cross
section
(\ref{2.1}) becomes \cite{MKDS, E} 
\be
\sigma_{\gamma^*_{L,T}p} (W^2, Q^2) = 
\frac{\alpha}{\pi} \sum_q Q^2_q Q^2 
\int dr^{\prime 2}_\bot K^2_{0,1} (r^\prime_\bot Q) 
\sigma_{(q \bar q)^{J=1}_{L,T} p} (r^{\prime 2}_\bot, W^2).
\label{2.3}
\ee
In the transition from (\ref{2.1}) to (\ref{2.3}), assuming massless quarks, we
inserted the explicit expressions for $\vert \psi_{L,T} (\vec r_\bot,
z (1-z), Q^2)\vert^2$ in terms of the modified Bessel functions $K_0
(r^\prime_\bot Q)$ and $K_1 (r^\prime_\bot Q)$, and $Q$ stands for 
$Q \equiv \sqrt{Q^2}$.The sum over the squared charges of the actively
contributing quarks is given by $\sum_q Q^2_q$, and the 
cross section $\sigma_{(q \bar q)^{J=1}_{L,T}} (r^\prime_\bot, W^2)$ is
related to the dipole cross section in (\ref{2.1}) by an appropriate
projection.

In terms of the $J=1$ projection, $\sigma_{(q \bar q)^{J=1}_{L,T}p}
(r^\prime_\bot, W^2)$, of the dipole cross section in (\ref{2.1}),
with $\vec l^{~\prime 2}_\bot = \vec l^{~ 2}_\bot/z (1-z)$, the
two-gluon-coupling structure of the dipole cross section in (\ref{2.2})
becomes \cite{E}
\be
\sigma_{(q \bar q)^{J=1}_{L,T}p} (r^\prime_\bot, W^2)  = 
\pi \int d \vec l^{~\prime 2}_\bot \bar \sigma_{(q \bar q)^{J=1}_{L,T} p}
(\vec l^{~\prime 2}_\bot , W^2) 
\cdot \left( 1 - \frac{\int d \vec l^{~\prime 2}_\bot 
\bar \sigma_{(q \bar q)^{J=1}_{L,T} p} (\vec l^{~\prime 2}_\bot, W^2) J_0
(l^\prime_\bot r^\prime_\bot)}{\int d \vec l^{~\prime 2}_\bot
\bar \sigma_{(q \bar q)^{J=1}_{L,T} p} (\vec l^{~\prime 2}_\bot, W^2)}
\right), 
\label{2.4}
\ee
where $J_0 (l^\prime_\bot r^\prime_\bot)$ denotes the Bessel function with
index $0$.

Two distinct $W$-dependent limits of the dipole cross section (\ref{2.4}) will
be relevant \cite{E} and important for the ensuing discussions. We assume
that the integrals in (\ref{2.4}) exist and are determined by a
$W$-dependent restricted range of $\vec l^{~\prime 2}_\bot < \vec
l^{~\prime_2}_{Max} (W^2)$, in which $\bar \sigma_{(q \vec q)^{J=1}_{L,T} p} 
(\vec l^{~\prime 2}_\bot, W^2)$ is appreciably different from zero, 
$\vec l^{~\prime 2}_{\bot~ Max} (W^2)$ increasing with increasing $W^2$.
The resulting cross section for a dipole of fixed transverse size, 
$r^\prime_\bot$,
strongly depends on the variation of the phase $l^\prime_\bot r^\prime_\bot$:
\begin{itemize}
\item[A)] For
\be
0 < l^\prime_\bot r^\prime_\bot < l^\prime_{\bot~Max} (W^2) r^\prime_\bot
\ll 1,
\label{2.5}
\ee
upon employing the expansion
\be
J_0 (l^\prime_\bot r^\prime_\bot) \cong 1 - \frac{1}{4} (l^\prime_\bot
r^\prime_\bot)^2 + \frac{1}{4^3} (l^\prime_\bot r^\prime_\bot)^4 + \cdots ,
\label{2.6}
\ee
we find a strong cancellation between the two additive contributions
to the $J=1$ dipole cross section (\ref{2.4}) which correspond to the first and
the second diagram in \break Fig.\,9. The dipole cross section (\ref{2.4}) 
becomes
proportional to $r^{\prime 2}_\bot$ (``color transparency'' 
limit \cite{Nikolaev})
\be
\sigma_{(q \bar q)^{J=1}_{L,T} p} (r^{\prime 2}_\bot, W^2) =
\frac{1}{4} r^{\prime 2}_\bot \sigma_L^{(\infty)} (W^2)
\Lambda^2_{sat} (W^2) 
\left\{ \matrix{ 1,\cr
\rho_W ,} \right. 
~~~~\left(r^{\prime 2}_\bot \ll
\frac{1}{l^{\prime 2}_{\bot~Max} (W^2)}\right),
\label{2.7}
\ee
where by definition the $W^2$-dependent scale $\Lambda^2_{sat} (W^2)$ reads
\be
\Lambda^2_{sat} (W^2) \equiv \frac{1}{\sigma^{(\infty)}_L(W^2)} \pi  
\int d \vec l^{~\prime 2}_\bot
\vec l^{~\prime 2}_\bot \bar \sigma_{(q \bar q)^{J=1}_L p} 
(\vec l^{~\prime 2}_\bot , W^2),
\label{2.8}
\ee
and $\sigma^{(\infty)}_L (W^2)$ is explicitly defined by (\ref{2.11}) below.
For vanishing size, $\vec r^{~\prime 2}_\bot$, the $q \bar q$ color dipole
(obviously) has a vanishing cross section. The expression for the factor
$\rho_W$ in (\ref{2.7}) is explicitly obtained by comparison of (\ref{2.7})
with (\ref{2.4}). According to (\ref{2.7}), and anticipating $\rho_W > 1$,
transversely polarized $(q \bar q)^{J=1}$ states interact with enhanced
transverse size \cite{Ku-Schi, E}
\be
\vec r^{~\prime 2}_\bot \rightarrow \rho_W \vec r^{~\prime 2}_\bot
\label{2.9}
\ee
relative to longitudinal ones. The ratio $\rho_W$ will be shown to be a
$W$-independent constant of definite magnitude, $\rho_W = \rho = 4/3$.
\item[B)] For the case of
\be
l^\prime_{\bot Max} (W^2) r^\prime_\bot \gg 1,
\label{2.10}
\ee
alternative to (\ref{2.5}), for any fixed value of $r^\prime_\bot$, rapid
oscillations of the Bessel function in (\ref{2.4}) lead to a vanishingly 
small contribution of the second term in (\ref{2.4}) thus implying
(``saturation'' limit)
\be
\hspace*{-0.3cm}
\lim_{r^{\prime 2} fixed \atop W^2 \to \infty} 
\sigma_{(q \bar q)^{J=1}_{L,T} p} (r_\bot^{~\prime 2}, W^2) 
= \pi
\int d \vec l^{~\prime 2}_\bot \bar \sigma_{(q \bar q)^{J=1}_{L,T} p} 
(\vec l^{~\prime 2}_\bot , W^2),
\left(r^{\prime 2}_\bot \gg
\frac{1}{l^{\prime 2}_{\bot~Max} (W^2)} \right).
\label{2.11}
\ee
The high-energy limit in (\ref{2.11}) of sufficiently large $W$ at fixed
dipole size $r^\prime_\bot$, according to (\ref{2.4}), coincides with the
limit of  sufficiently large $r^\prime_\bot$ at fixed $W$ i.e.
\be
\lim_{r^{\prime 2} fixed \atop W^2 \to \infty} 
\sigma_{(q \bar q)^{J=1}_{L,T} p} (r_\bot^{~\prime 2}, W^2) 
=
\lim_{r^{\prime 2}_\bot \to \infty \atop W^2 = const} 
\sigma_{(q \bar q)^{J=1}_{L,T}p} (r^{\prime 2}_\bot, W^2) = 
\sigma_{L,T}^{(\infty)} (W^2).
\label{2.12}
\ee
At fixed dipole size, at sufficiently large energy, we arrive at the
large-dipole-size limit of the (at most weakly $W$-dependent) hadronic
cross section $\sigma^{(\infty)}_{L,T} (W^2) \simeq \sigma^{(\infty)}$.
\end{itemize}

The photoabsorption cross section in (\ref{2.3}), due to the strong 
decrease of the modified Bessel functions $K_{0,1} (r^\prime_\bot~Q)$ with
increasing argument $r^\prime_\bot Q$, is strongly dominated and actually 
determined at any fixed value of $Q^2$ by values of $r^{\prime~2}_\bot$
such that $r^{\prime 2}_{\bot} Q^2 < 1$. Whether color transparency of the
dipole cross section according to (\ref{2.5}) and (\ref{2.7}) or, 
alternatively, saturation
according to (\ref{2.10}) and
(\ref{2.11}) is relevant for $\sigma_{\gamma^*_{L,T}p} (W^2,Q^2)$
in (\ref{2.3}) depends on whether $Q^2 \gg \Lambda^2_{sat} (W^2)$ or 
$Q^2 \ll \Lambda^2_{sat} (W^2)$ is realized at a specific value of $W$.
Upon substitution of (\ref{2.7}) and, alternatively, of (\ref{2.11}) into
(\ref{2.3}), for $\sigma_{\gamma^*p} (W^2,Q^2) = \sigma_{\gamma^*_L p}
(W^2,Q^2) + \sigma_{\gamma^*_Tp} (W^2,Q^2)$ one finds \cite{DIFF2000, E}
\be
\sigma_{\gamma^*p} (W^2, Q^2)  =  \sigma_{\gamma^*p} (\eta
(W^2, Q^2)) = \frac{\alpha}{\pi} \sum_q Q^2_q  \cdot
\left\{ \matrix{
\frac{1}{6} ( 1 +2 \rho ) \sigma^{(\infty)}  \frac{1}{\eta(W^2, Q^2)},
~~ (\eta (W^2,Q^2) \gg 1), \cr
\sigma^{(\infty)}
\ln \frac{1}{\eta (W^2, Q^2)},~~~~~~~~~~~(\eta (W^2, Q^2) \ll 1), } \right.
\label{2.13}
\ee
in agreement with the experimental result (\ref{1.4}) shown in Fig.\,1. 
In (\ref{2.13}), we have introduced the low-x scaling variable \cite{DIFF2000}
\be
\eta (W^2, Q^2) = \frac{Q^2 + m^2_0}{\Lambda^2_{sat} (W^2)},
\label{2.14}
\ee
and anticipated $\rho_W = {\rm const.} = \rho$. In (\ref{2.14}), via
quark-hadron duality \cite{Sakurai, ST}, we introduced the 
lower bound $m^2_0 \lsim ~ m^2_\rho$ on the masses of the
$q \bar q$ fluctuations
$M^2_{q \bar q} \ge m^2_0$, only relevant in the limit of $Q^2 \to 0$.

From the above derivation, leading to (\ref{2.13}), it has become clear
that DIS at low x in terms of the Compton-forward-scattering amplitude
proceeds via two different reaction channels.
They correspond to the first and the second diagram in Fig.\,9. For
sufficiently large $Q^2 \gg \Lambda^2_{sat} (W^2)$ both channels are
open, resulting in strong cancellation between them. With
decreasing $Q^2$ at fixed $W^2$, or with increasing $W^2$ at fixed $Q^2$,
for $Q^2 \ll \Lambda^2_{sat} (W^2)$, the second channel becomes closed,
no cancellation any more. Only the first channel remains open,
implying that the proportionality of the photoabsorption cross section
to $\Lambda^2_{sat} (W^2)$ turns into the (soft) energy dependence
proportional to $\ln \Lambda^2_{sat} (W^2)$, compare (\ref{2.13}).

The longitudinal-to-transverse ratio of the photoabsorption cross sections
$\sigma_{\gamma^*_L p} (W^2,Q^2)$ and $\sigma_{\gamma^*_T p} (W^2,Q^2)$ at
large $Q^2 \gg \Lambda^2_{sat} (W^2)$ according to (\ref{2.13}) is given by
\be
R(W^2,Q^2)_{Q^2 \gg \Lambda^2_{sat} (W^2)}  = 
\frac{\sigma_{\gamma^*_L} (W^2,Q^2)}{\sigma_{\gamma^*_T} (W^2,Q^2)} 
\bigg|_{Q^2 \gg \Lambda^2_{sat} (W^2)} = \frac{1}{2 \rho}. 
\label{2.15}
\ee
The factor 2 in (\ref{2.15}) originates from the difference in the
photon wave functions in (\ref{2.3}). The interaction with enhanced
transverse size of $(q \bar q)^{J=1}_T$ states relative to
$(q \bar q)^{J=1}_L$ states, $\rho_W$ in (\ref{2.9}), is a consequence
of the ratio of the average transverse momenta of the quark (antiquark)
in the $(q \bar q)^{J=1}_T$ state relative to the quark (antiquark) in
the $(q \bar q)^{J=1}_L$ state. Upon applying the uncertainty principle, one
obtains \cite{Ku-Schi, E}
\be
\rho_W = \rho = \frac{4}{3}.
\label{2.16}
\ee
The longitudinal structure function, with (\ref{2.15}) and (\ref{2.16}), 
at large $Q^2$
is related to the transverse one via
\be
\hspace*{-0.7cm}F_L (x,Q^2) = \frac{1}{1+2 \rho} F_2 (x,Q^2) = 
0.27 F_2 (x,Q^2).
\label{2.17}
\ee 
The result is consistent with the experimental data, compare Fig.\,10. In
Fig.\,10, in addition to the experimental results on $F_L$ from the H1
and ZEUS collaborations, we also show the result on 
$(F_L/F_2)/(F_L/F_2)\vert_{bound}$ from a very detailed analysis of the
experimental data by the authors of ref. \cite{Nachtmann}. Since the
value of $(F_L/F_2)\vert_{bound} = 0.27$ from ref. \cite{Nachtmann}
(accidentally) numerically coincides with our prediction (\ref{2.17}),
the analysis of ref. \cite{Nachtmann} provides additional confirmation,
within errors, of the prediction (\ref{2.17}).

\begin{figure}[h]
\begin{center}
\includegraphics[scale=.3]{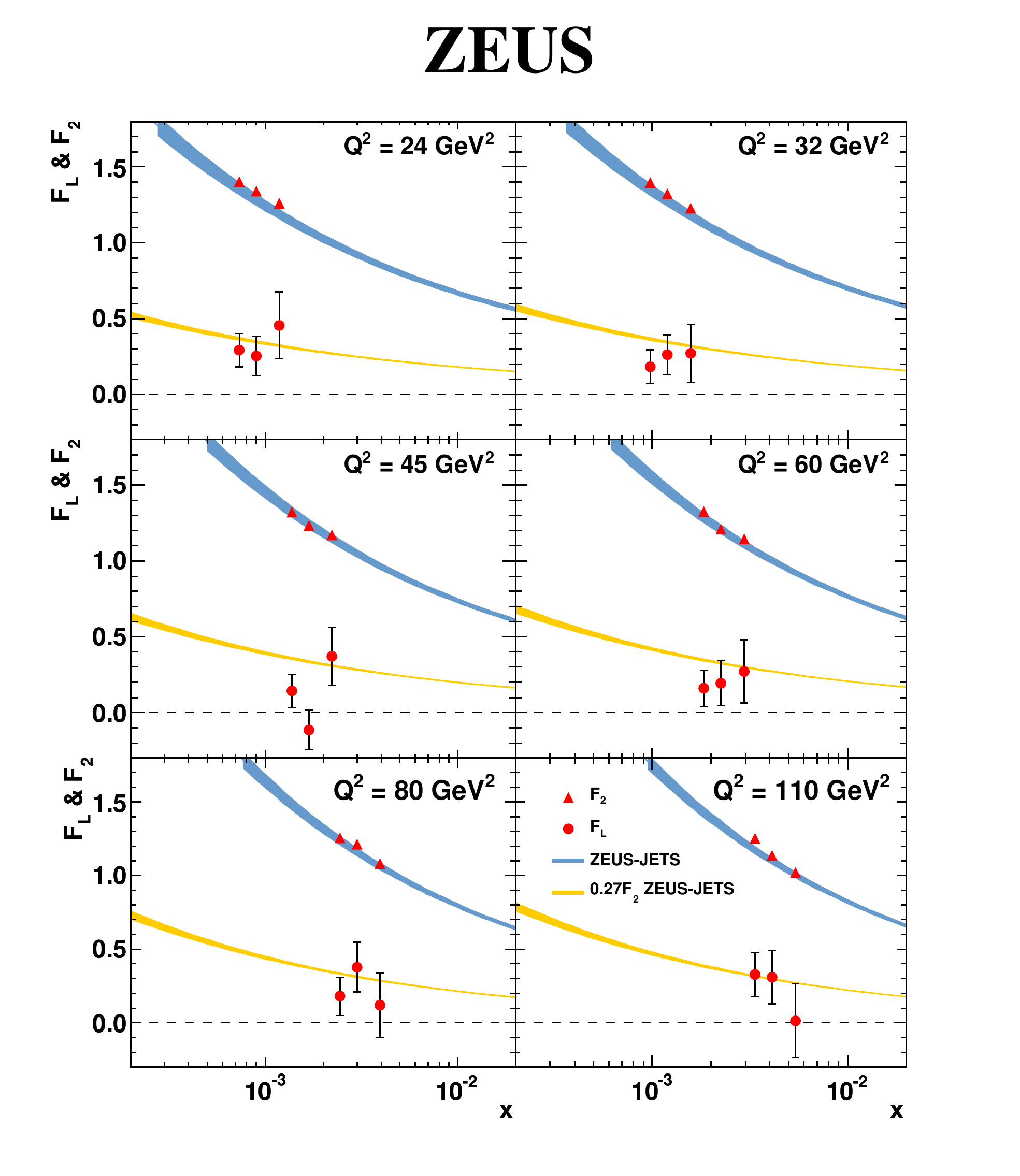}
\includegraphics[scale=.3]{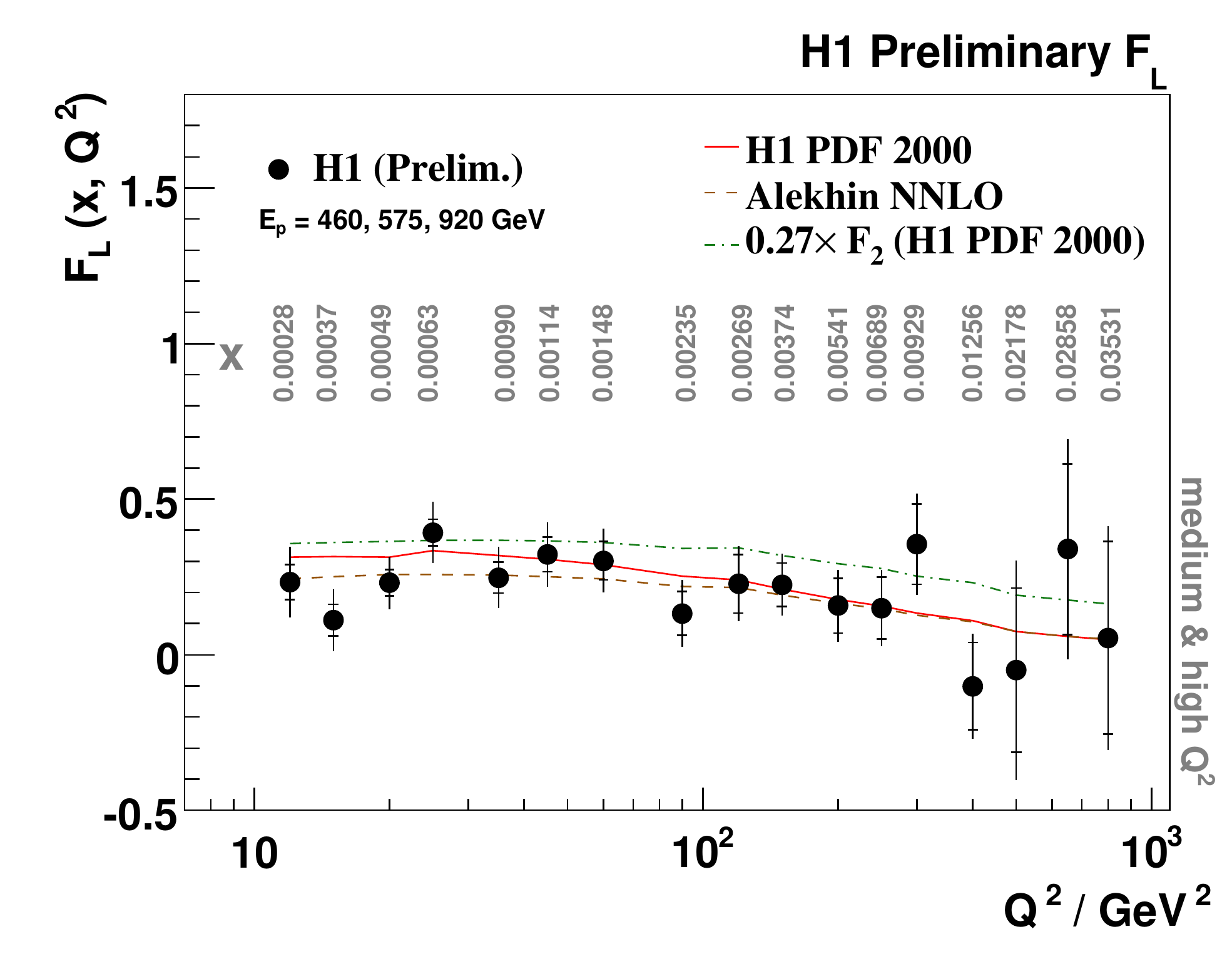}
\includegraphics[scale=.6]{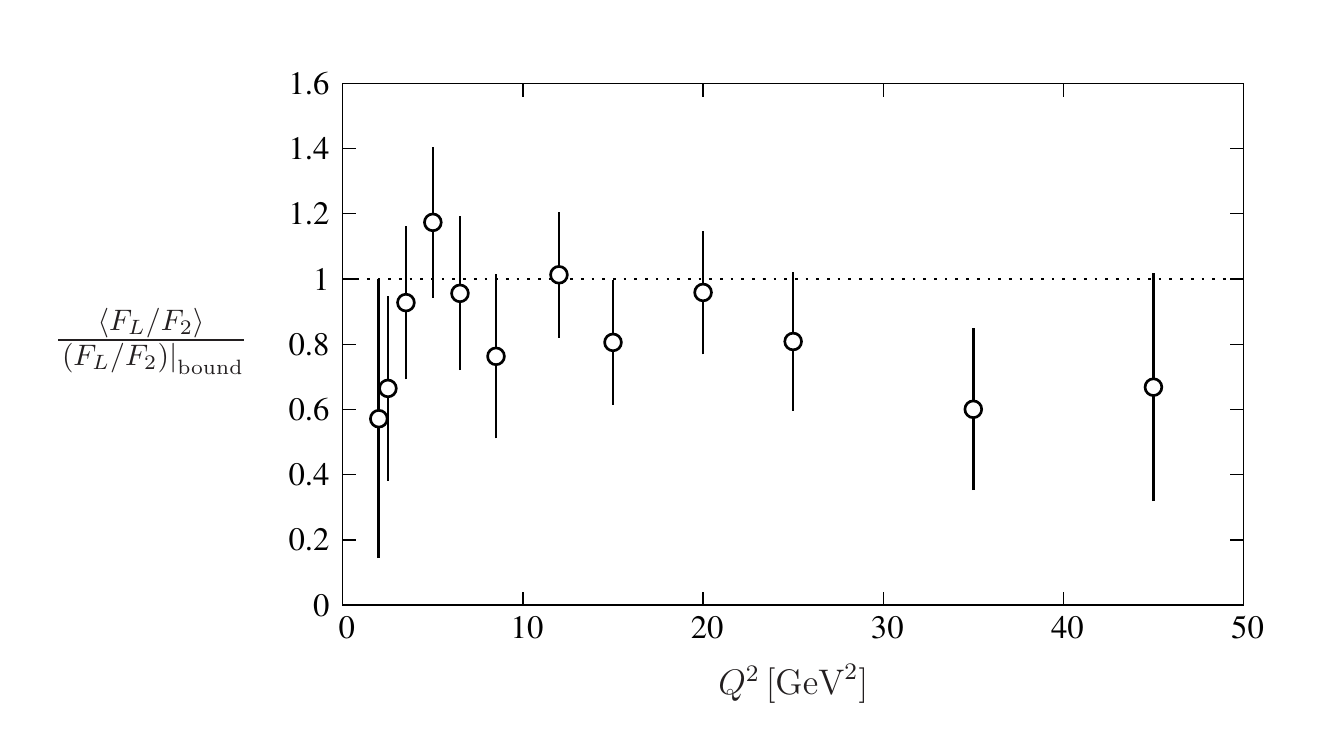}
\vspace*{-0.3cm}
\caption{The prediction (\ref{2.17}) compared with the experimental
data for $F_L$ and $F_2$.}
\end{center}
\vspace*{-0.3cm}
\end{figure}

The $W$-dependence in (\ref{2.1}) and (\ref{2.3}) of the dipole cross
section, combined with the $1/Q^2$ dependence in (\ref{2.13}) at
large $Q^2$, implies that the structure function $F_2 (x,Q^2) \simeq
(Q^2/4 \pi^2 \alpha) (\sigma_{\gamma^*_L p} (W^2, Q^2) + 
\sigma_{\gamma^*_T p} (W^2,Q^2))$ becomes a function of the single variable
$W^2$. The experimental data in Fig.\,11, in the relevant range of
$x \simeq Q^2/W^2 < 0.1$, approximately corresponding to $1/W^2 \le
10^{-3}$, indeed show a single line 
\cite{E} \footnote{The representation of the experimental data in
Fig.\,11 was kindly prepared by Prabhdeep Kaur.} consistent with the
representation of $\sigma_{\gamma^*p} (\eta(W^2,Q^2))$ in Fig. 1.
An eye-ball fit to the
data in Fig.\,11 yields
\be
F_2 (W^2) = f_2 \cdot \left( \frac{W^2}{1{\rm GeV}^2} \right)^{C_2 = 0.29}  
\label{2.18}
\ee
with $f_2 = 0.063$. We shall come back to representation (\ref{2.18}) of
$F_2(W^2)$ below. For comparison, in Fig.\,11, we also show $F_2(x,Q^2)$
as a function of the Bjorken variable $x$.

\begin{figure}[h!]
\begin{center}
\vspace*{-0.5cm}
\includegraphics[scale=.5]{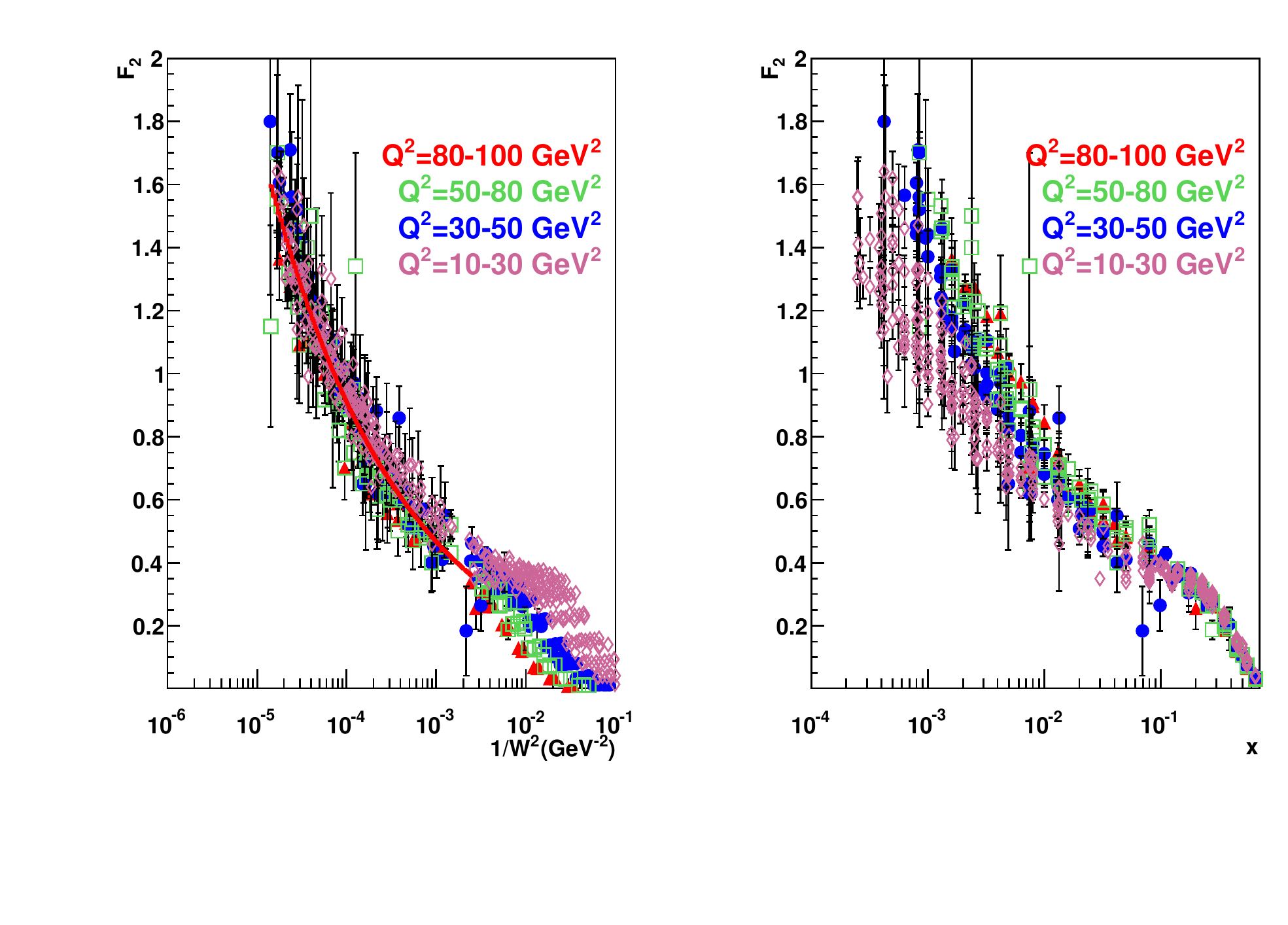}
\vspace*{-1cm}
\caption{The structure function $F_2 (x, Q^2)$ as a function of
$1/W^2$ and as a function of $x$}
\vspace*{-0.5cm}
\end{center}
\end{figure}


We turn to the limit of $\eta (W^2,Q^2) \ll 1$, or
$W^2 \to \infty$ at $Q^2$ fixed in (\ref{2.13}), (\ref{1.4}) and (\ref{1.6}). 
The convergence to this limit is extremely slow. Compare
Fig.\,12, where this limit is shown in terms of the structure function
$F_2 (x,Q^2)$,
\be
\lim\limits_{{W^2\rightarrow\infty}\atop{Q^2 {\rm fixed}}}
\frac{F_2 (x\cong Q^2 / W^2, Q^2)}{\sigma_{\gamma p} (W^2)} =
\frac{Q^2}{4\pi^2\alpha} .
\label{2.20}
\ee 
\begin{figure}[h!]
\begin{center}
\vspace*{-0.8cm}
\includegraphics[scale=.35]{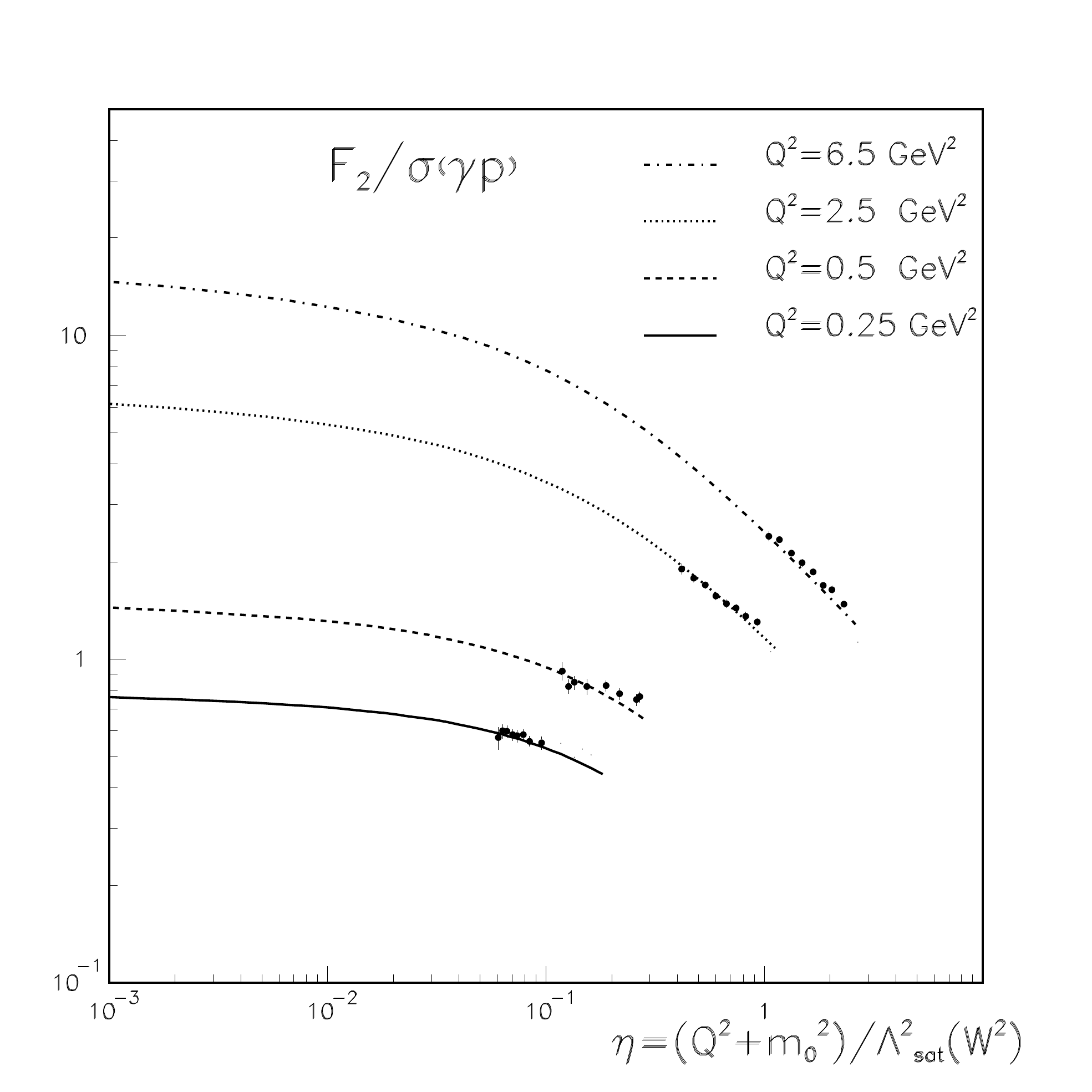}
\vspace*{-0.3cm}
\caption{The approach to saturation}
\vspace*{-0.5cm}
\end{center}
\end{figure}

The theoretical curve in Fig.\,12 is due to the concrete ansatz for the dipole
cross section in Section 4 that interpolates between the regions of
$\eta (W^2,Q^2) \gg 1$ and $\eta (W^2,Q^2) \ll 1$.

An approach to a $Q^2$-independent limit for $W^2 \to \infty$ at
fixed $Q^2$ was recently independently observed by Caldwell \cite{CAL}
based on a purely empirical fit to the experimental data given by
\be
\sigma_{\gamma^*p} (W^2,Q^2) = \sigma_0 (Q^2)
\left( \frac{1}{2M_p} \frac{W^2}{Q^2} \right)^{\lambda_{eff} (Q^2)}.
\label{2.21}
\ee
The straight lines in Fig.\,13 meet at a value of $W^2$ approximately
given by $W^2 \simeq 10^9 Q^2$, consistent with the above conclusion 
from the CDP.
\begin{figure}[h!]
\begin{center}
\vspace*{-3.4cm}
\includegraphics[scale=.35]{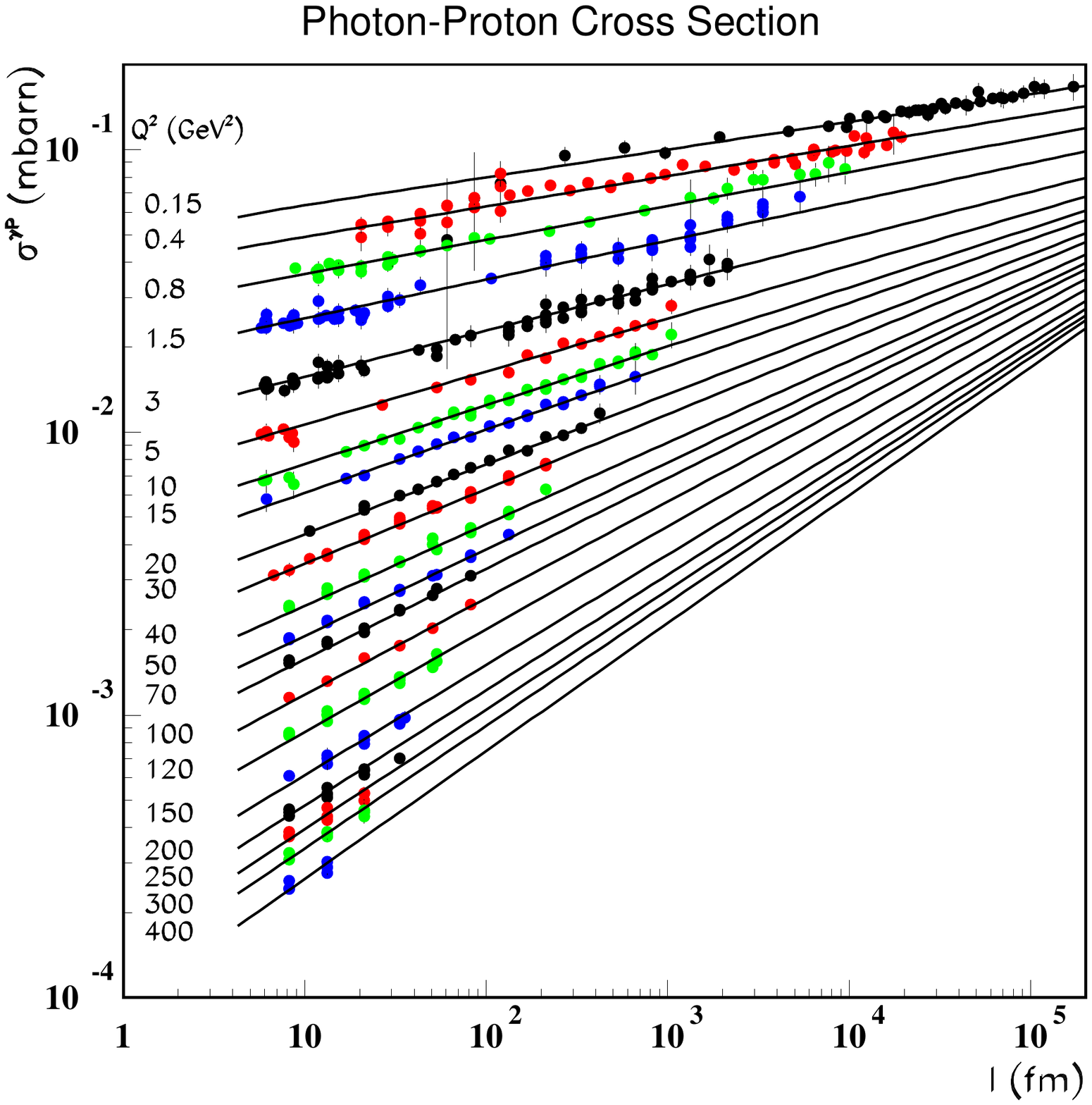}
\caption{The Caldwell fit (\ref{2.21})\cite{CAL}}
\end{center}
\vspace*{-1cm}
\end{figure}

The results from the above general analysis of the CDP lead to the
simple structure of the $(Q^2, W^2)$ plane shown in Fig.\,2. 


The $(Q^2,W^2)$ plane is subdivided into (only) two regions as a consequence of
the two interaction channels corresponding to the diagrams in Fig.\,9.
For $\eta (W^2,Q^2) > 1$, both interaction channels, channels 1
and 2, are open, implying color transparency of the $(q \bar q)p$
dipole interaction and strong destructive interference in the photoabsorption
cross section for $\gamma^*p \to {\rm hadrons}$. For
$\eta(W^2,Q^2) \ll 1$, channel 2 becomes closed. The lack of
cancellation leads to a saturation of the cross section being
determined by the dipole interaction solely via channel 1. The
virtual-photon-nucleon cross section approaches a $Q^2$-independent
saturation limit that coincides with $Q^2 = 0$ photoproduction.

\section{The CDP, the gluon distribution function and evolution}
\renewcommand{\theequation}{\arabic{section}.\arabic{equation}}
\setcounter{equation}{0}

The CDP of DIS corresponds\footnote{With respect to this Section,
compare also ref. \cite{23a}} to the low $x$ approximation of the
pQCD-improved parton model in which the interaction of the (virtual)
photon occurs by interaction with the quark-antiquark sea in the 
proton via $\gamma^* gluon \to q \bar q$ fusion.

The longitudinal structure function in this approximation, for a wide
range of different gluon distributions, becomes proportional to the
gluon density at a rescaled value $x/\xi_L$ \cite{Martin}
\be 
F_L ( \xi_L x, Q^2) = \frac{\alpha_s (Q^2)}{3\pi} \sum_q Q^2_q G (x, Q^2),
\label{3.1}
\ee
where $G(x,Q^2) \equiv xg (x,Q^2)$ and $g(x,Q^2)$ stands for the 
gluon-distribution 
function. The rescaling factor has the preferred value of
$\xi_L \simeq 0.40$.

The structure function $F_2 (x,Q^2)$ at low $x$ in this approximation
is proportional to the sea-quark distribution, and again for a wide
range of different gluon distributions, the evolution of the structure
function $F_2 (x,Q^2)$ with $Q^2$ is determined by \cite{Lipatov, Prytz}
\be
\frac{\partial F_2 (\xi_2 x , Q^2)}{\partial \ln Q^2} = \frac{\alpha_s
  (Q^2)}{3\pi} \sum_q Q^2_q G  (x ,Q^2),
\label{3.2}
\ee
where the preferred value of $\xi_2$ is given by $\xi_2 \simeq 0.50.$

According to the CDP, compare (\ref{2.13}) and (\ref{2.18}), 
and supported by the 
experimental data in Fig.\,11,
the structure function $F_2 (x,Q^2)$ for sufficiently large $Q^2$ becomes
a function of the single variable $W^2$,
\be
F_2 (x, Q^2) = F_2 (W^2 = \frac{Q^2}{x}) . 
\label{3.3} 
\ee
Employing the proportionality of $F_L(x,Q^2) = (1/(1+2 \rho_W)) F_2(x,Q^2)$
from (\ref{2.7}) and (\ref{2.17}), and combining (\ref{3.1}) and 
(\ref{3.2}), upon inserting (\ref{3.3}) the evolution equation becomes
\be
(2 \rho_W + 1) \frac{\partial}{\partial \ln W^2} F_2 \left( \frac{\xi_L}{\xi_2}
  W^2 \right) = F_2 (W^2). 
\label{3.4}
 \ee
A potential dependence of $\rho_W$ on the energy $W$ is allowed in 
(\ref{3.4}).

We specify $F_2 (W^2)$ by adopting the power law \cite{E}
\be
F_2 (W^2) \sim (W^2)^{C_2} = \left( \frac{Q^2}{x} \right)^{C_2} . 
\label{3.5}
\ee
A power law in $(1/x)^\lambda$ with $\lambda = const.$ occurs e.g. in the
``hard Pomeron solution'' \cite{Adel} of DGLAP evolution as well
as in the ``hard Pomeron'' part of Regge phenomenology with
$(1/x)^{\epsilon_0}$ and $\epsilon_0 \simeq 0.43$ from a fit
\cite{Dom}. The CDP in (\ref{3.5}) is more specific, however,
since the $W$ dependence of $F_2(W^2)$ implies that the $x$ dependence
and the $Q^2$ dependence are intimately related to each other.

Substitution of (\ref{3.5}) into (\ref{3.4}) implies the constraint
\cite{N9, E}
\be
(2\rho_W + 1) C_2 \left( \frac{\xi_L}{\xi_2} \right)^{C_2} = 1 . 
\label{3.6}
\ee
If, and only if $\rho_W = \rho = {\rm const.}$, also the exponent
is constant, $C_2 = {\rm const.}$ With the CDP result of $\rho =
4/3$ from (\ref{2.16}), we obtain the unique value of
\be
C_2 = \frac{1}{2\rho + 1} \left( \frac{\xi_2}{\xi_L} \right)^{C_2} = 
0.29,
\label{3.7}
\ee 
where $\rho = 4/3$ and $\xi_2/\xi_L = 1.25$ was inserted. The result
for $C_2 = 0.29$ is fairly insensitive under variation of $\xi_2/\xi_L$.
For $1 \le \xi_2/\xi_L \le 1.5$, one obtains $0.27 \le C_2 \le 0.31$.
The value of $C_2 = 0.29$ is consistent with the experimental data,
compare (\ref{2.18}) and the corresponding theoretical curve in Fig.\,11.

Imposing consistency between the CDP and the pQCD-improved parton
model, we thus arrived at the prediction of a definite value of
$C_2 = 0.29$ that coincides with the experimental findings in Fig.\,11.

\begin{figure}[h!]
\begin{center}
\vspace*{-2.3cm}
\includegraphics[scale=.4]{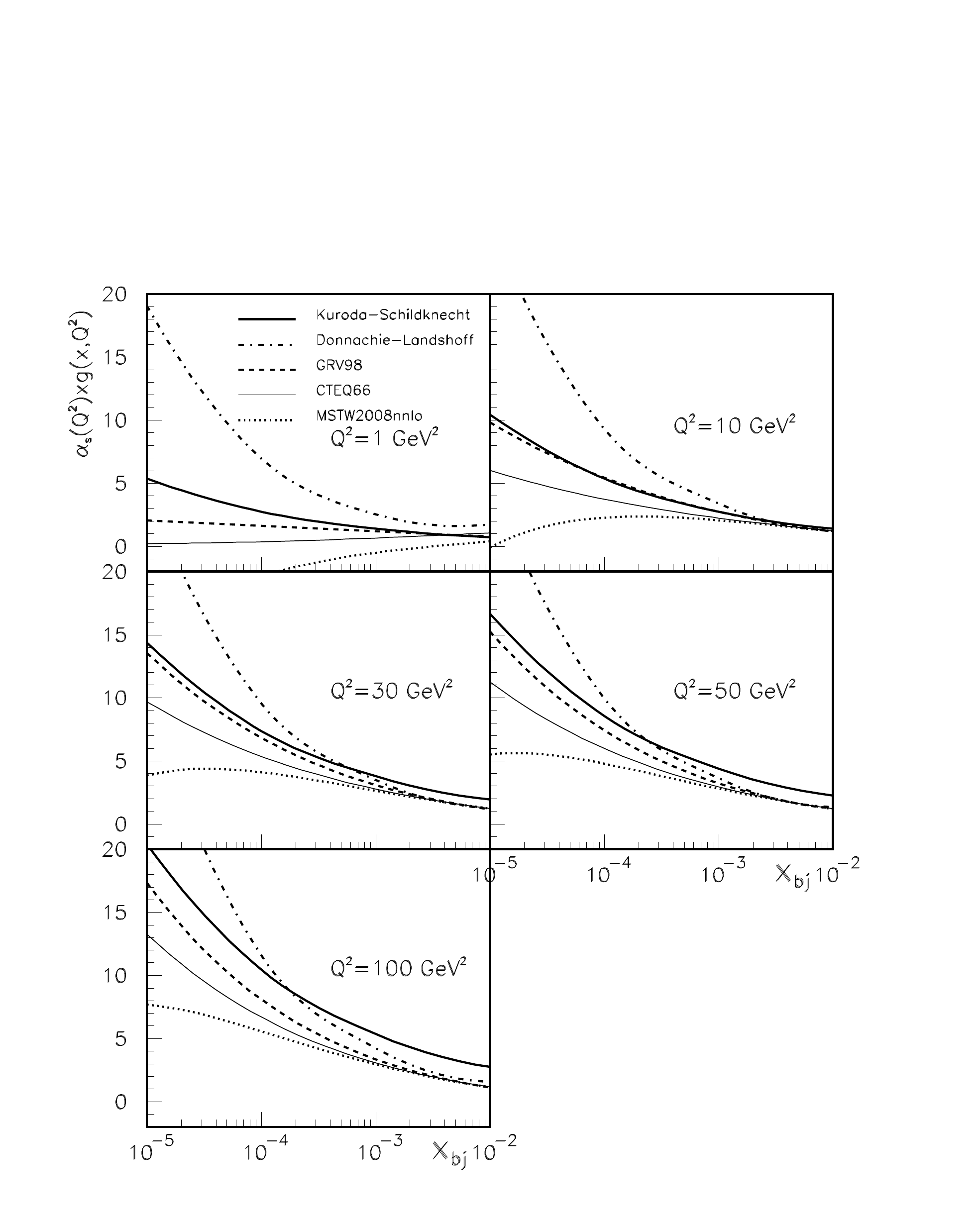}
\vspace*{-0.8cm}
\caption{The gluon distribution function (\ref{3.8}) compared with
the results from the hard Pomeron part of a Regge fit to $F_2 (x,Q^2)$
and from the fits GRV \cite{GRV} CTEQ \cite{CTEQ} and \cite{MSTW}.}
\end{center}
\vspace*{-0.8cm}
\end{figure}

The underlying gluon distribution function can now be deduced from
(\ref{3.1}) by expressing the longitudinal structure function in terms
of $F_2 (x,Q^2)$ according to (\ref{2.17}) and inserting the power law
(\ref{2.18}),
\be
\alpha_s (Q^2) G (x, Q^2)  
=\frac{3\pi}{\sum_q Q^2_q (2\rho + 1)} 
\frac{f_2}{\xi_L^{C_2 = 0.29}} \left(
  \frac{W^2}{1 {\rm GeV}^2} \right)^{C_2 = 0.29}. 
\label{3.8}
\ee
With $\rho = 4/3$ from the CDP, compare (\ref{2.16}), and $\sum_q Q^2_q =
10/9$,
the result (\ref{3.8}) contains the
single fitted parameter $f_2 = 0.063$ from (\ref{2.18}). Inserting
$W^2 = Q^2/x$ on the right-hand side in (\ref{3.8}), 
we obtain the gluon distribution as
a function of $x$ and $Q^2$.

Using the next-to-leading order expression for $\alpha_s(Q^2)$, in
Fig.\,14 \cite{E}, we compare the gluon distribution (\ref{3.8}) with
various gluon distributions obtained in sophisticated global fits 
\cite{GRV, CTEQ, MSTW}to the 
experimental data. The consistency of our simple one-free-parameter
extraction of the gluon distribution in Fig.\,14 according to
(\ref{3.8}) may indicate that the gluon distribution is less sensitively
dependent on the details of the $ggpp$ structure than usually assumed,
or elaborated upon and employed in the global fits to the experimental data.

\section{Specific ansatz for the dipole cross section and comparison with
  experiment}
\renewcommand{\theequation}{\arabic{section}.\arabic{equation}}
\setcounter{equation}{0}

Any specific ansatz for the dipole cross section 
has to interpolate between the color-transparency
region of $\eta (W^2,Q^2) \gg 1$, where $\sigma_{\gamma^*p} (\eta (W^2,Q^2))
\sim 1/\eta (W^2,Q^2)$, and the saturation region of 
$\eta (W^2,Q^2) \ll 1$, where
$\sigma_{\gamma^*p} (\eta(W^2,Q^2)) \sim \ln \left( 1/\eta (W^2,Q^2)\right)$,
compare (\ref{2.13}). For the explicit expressions for the ansatz for the
dipole cross section, we refer to refs. \cite{DIFF2000} and \cite{E}
\footnote{For deeply virtual Compton scattering and vector
meson production in the CDP compare ref. \cite{Kuroda}}.
We only note that the saturation scale, $\Lambda^2_{sat} (W^2) \sim
(W^2)^{C_2}$ in the HERA energy range approximately varies between
$2 GeV^2 \le \Lambda^2_{sat} (W^2) \le 7 GeV^2$, and
restrict ourselves to presenting a comparison with the experimental data.

\begin{figure}[h!]
\begin{center}
\vspace*{-0.1cm}
\includegraphics[scale=.5]{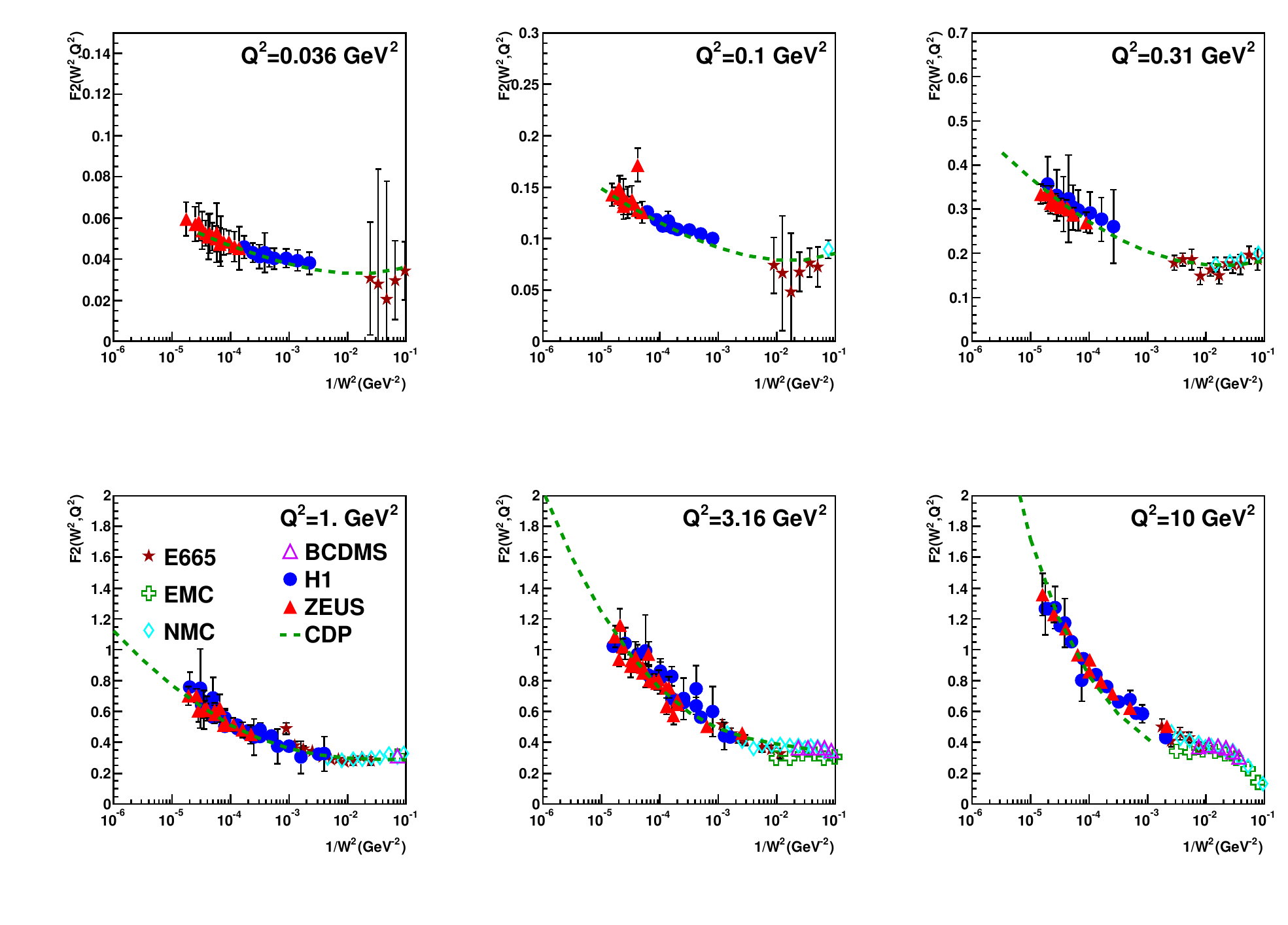}
\vspace*{-0.5cm}
\caption{The predictions \cite{E} from the CDP for the structure function
$F_2 (W^2,Q^2)$ compared with the experimental data for $0.036 \le Q^2 \le
10 GeV^2$.}
\end{center}
\end{figure}
\begin{figure}[h!]
\begin{center}
\vspace*{-0.5cm}
\includegraphics[scale=.5]{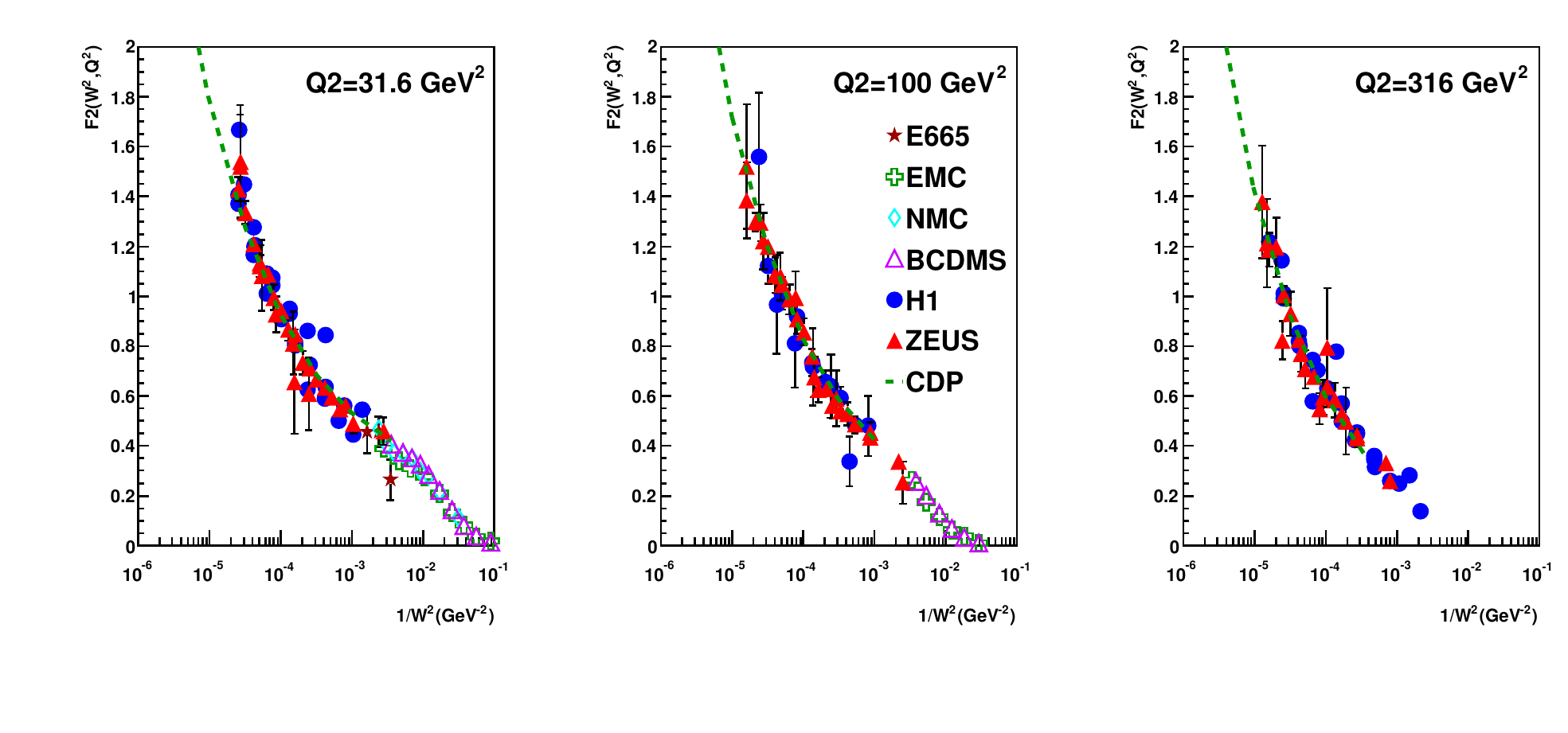}
\vspace*{-0.7cm}
\caption{As in Fig.\,15, but for $31.6 GeV^2 \le Q^2 \le 316 GeV^2$.}
\vspace*{-0.5cm}
\end{center}
\end{figure}

The theoretical results from the CDP in Figs. 15 and 16 show agreement
with the experimental data\footnote{We thank Prabhdeep Kaur for providing
the plots of the experimental data in Figs 15 to 18.} 
for $F_2 (W^2,Q^2)$ over the full relevant
region of $0.036~GeV^2 \le Q^2 \le 316 GeV^2$.

In Figs.\,17 and 18, in addition to the theoretical results from the CDP,
for comparison, we also show the results from the pQCD improved parton
model based on the gluon distribution function (\ref{3.8}) shown in
Fig.\,14.

Explicitly, by returning to (\ref{2.18}) and inverting (\ref{3.8}), we
reinterprete (\ref{3.8}) as a prediction of $F_2 (W^2 = Q^2/x)$
from the (previously determined)
gluon distribution according to 
\be
F_2 (W^2 = Q^2/x) = 
\frac{(2 \rho + 1) \sum_q Q^2_q}{3 \pi} 
\xi^{C_2=0.29}_L \alpha_s (Q^2) G(x,Q^2),
\label{4.1}
\ee
where $f_2 = 0.063$ and $\xi_L = 0.40$. In Figs. 17 and 18, we see the 
expected consistency of the
pQCD prediction (\ref{4.1}) with the experimental data 
and the CDP in the 
relevant range of $10 GeV^2 \le Q^2 \le 100 GeV^2$. For $Q^2 < 10 GeV^2$,
with increasing $W$, gradually saturation sets in implying a breakdown of
the pQCD proportionality (\ref{4.1}) of the proton structure function to
the gluon distribution. 

\begin{figure}[h!]
\begin{center}
\includegraphics[scale=.5]{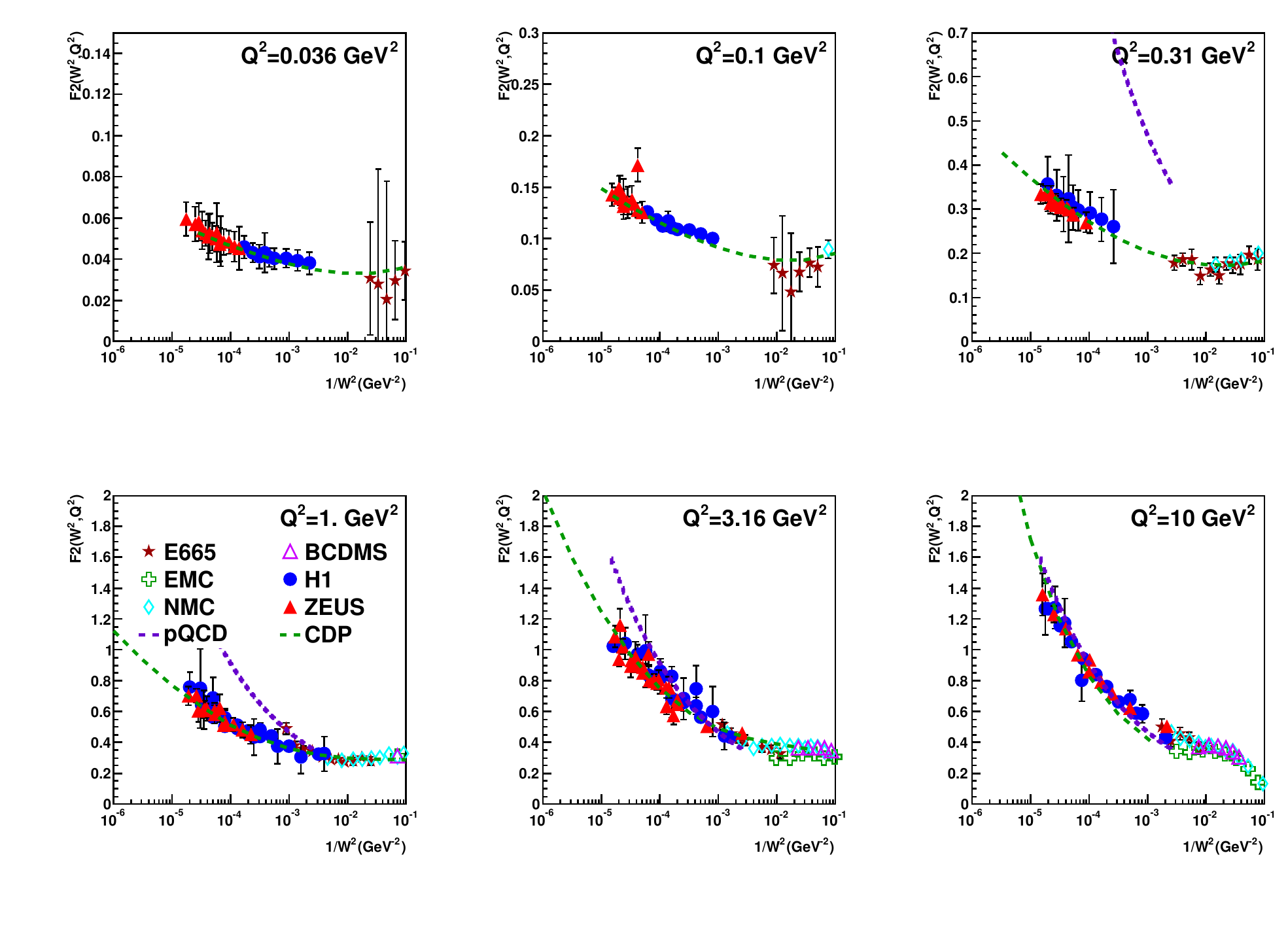}
\vspace*{-0.5cm}
\caption{In addition to the prediction from the CDP, also the pQCD prediction
(\ref{4.1}) based on the gluon distribution (\ref{3.8}) is compared with
the experimental data for $F_2 (W^2,Q^2)$.}
\end{center}
\end{figure}

\begin{figure}[h!]
\begin{center}
\vspace*{-0.5cm}
\includegraphics[scale=.5]{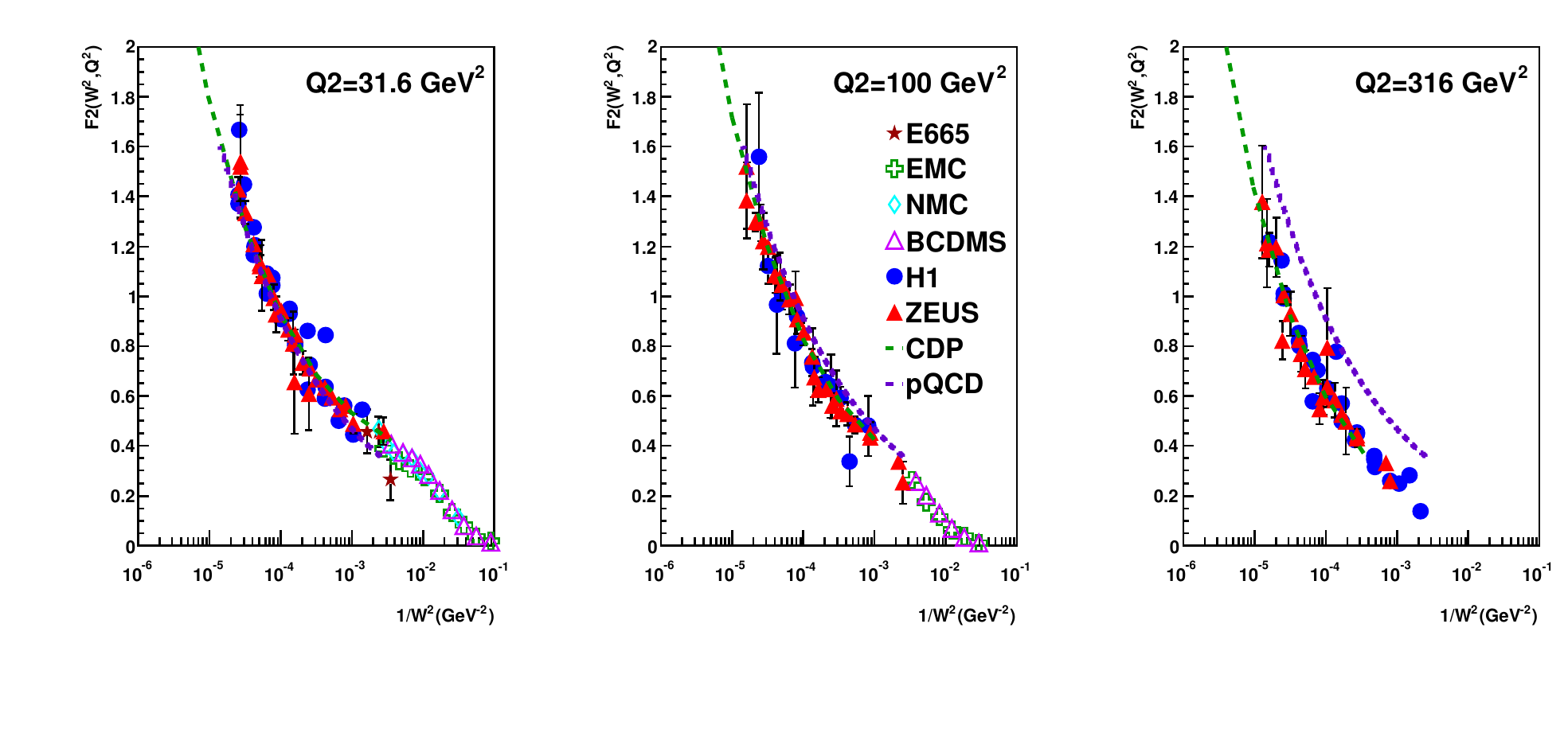}
\vspace*{-0.7cm}
\caption{As in Fig.\,17, but for $31.6 GeV^2 \le Q^2 \le 316 GeV^2$.}
\vspace*{-0.4cm}
\end{center}
\end{figure}

The pQCD prediction
for $Q^2 = 316 GeV^2$ lies above the experimental data. This is due to the
breakdown of the simple form for $F_2 (W^2)$ in (\ref{2.18}) which is 
used to extract the gluon distribution. Employing the full CDP result
would lead to an appropriate decrease of the gluon distribution with
increasing $Q^2$ for $Q^2 > 100 GeV^2$.

The proton structure function of the CDP, according to (\ref{2.13}),
in the region of $\eta (W^2,Q^2) 
< 1$, starts to depend logarithmically on the saturation scale,
$\Lambda^2_{sat} (W^2) \sim (W^2)^{C_2}$, and with $\alpha_s (Q^2) G(x,Q^2)
\sim (W^2)^{C_2}$ from (\ref{3.8}) and (\ref{4.1}), it depends
logarithmically on the gluon distribution,
\be
F_2 (W^2,Q^2) \sim Q^2 \sigma_L^{(\infty)} \ln 
\frac{\Lambda^2_{sat}
(W^2 = Q^2/x)}{Q^2 + m^2_0} 
\sim  Q^2 \sigma_L^{(\infty)} \ln \frac{\alpha_s (Q^2) G (x,Q^2)}
{\sigma_L^{(\infty)} (Q^2 + m^2_0)}, \left( Q^2 \ll \Lambda^2_{sat}
  (W^2)\right).
\label{4.2}
\ee
The smooth transition from the color transparency region to the saturation
region does not correspond to a change of the functional form of 
the $W$-dependent gluon
distribution, $\alpha_s (Q^2) G(x,Q^2) \sim \Lambda^2_{sat}(W^2)$, 
but occurs via transition from
the proportionality (\ref{4.1}) to the logarithmic  dependence
(\ref{4.2}) of $F_2 (W^2, Q^2)$
on the gluon distribution function. It is the same gluon
distribution that is relevant in the region of color transparency
$\eta (W^2,Q^2) \gg 1$ and in the saturation region, $\eta (W^2, Q^2) \ll 1$,
only the functional
dependence of the proton structure function $F_2 (x,Q^2)$
on the gluon distribution has changed. We disagree
with the frequently expressed opinion (compare e.g. ref. \cite{F} and 
the list of references given therein) that the mere
existence of the saturation scale and of scaling like in Fig.\,1
suggests and even requires a modification based on non-linear evolution
of the gluon distribution determined in the pQCD domain of 
$\eta (W^2,Q^2) > 1$,
when passing to the saturation region of $\eta (W^2,Q^2) < 1$.

\begin{figure}[h]
\begin{center}
\vspace*{-0.2cm}
\includegraphics[scale=.25]{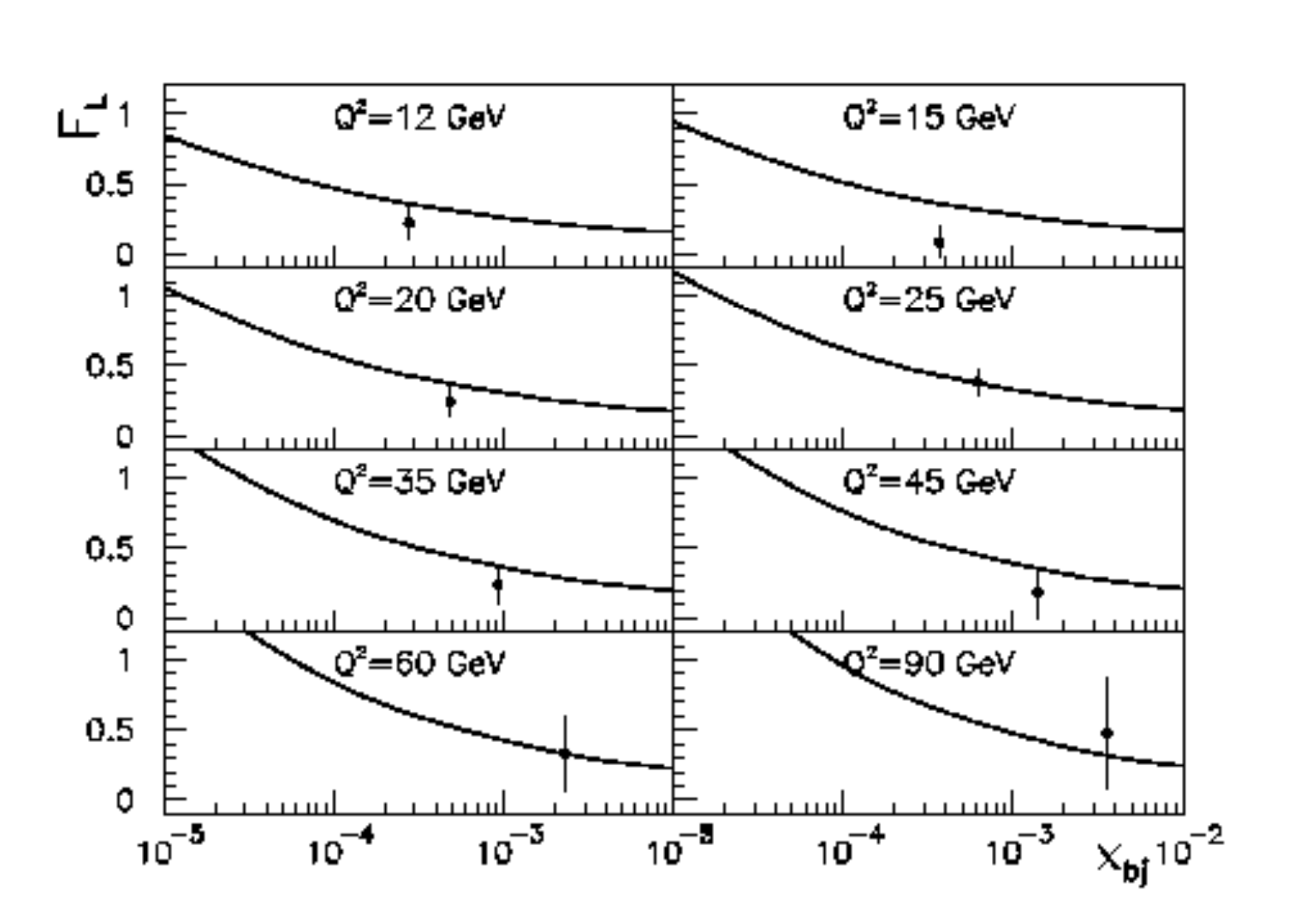}
\includegraphics[scale=.25]{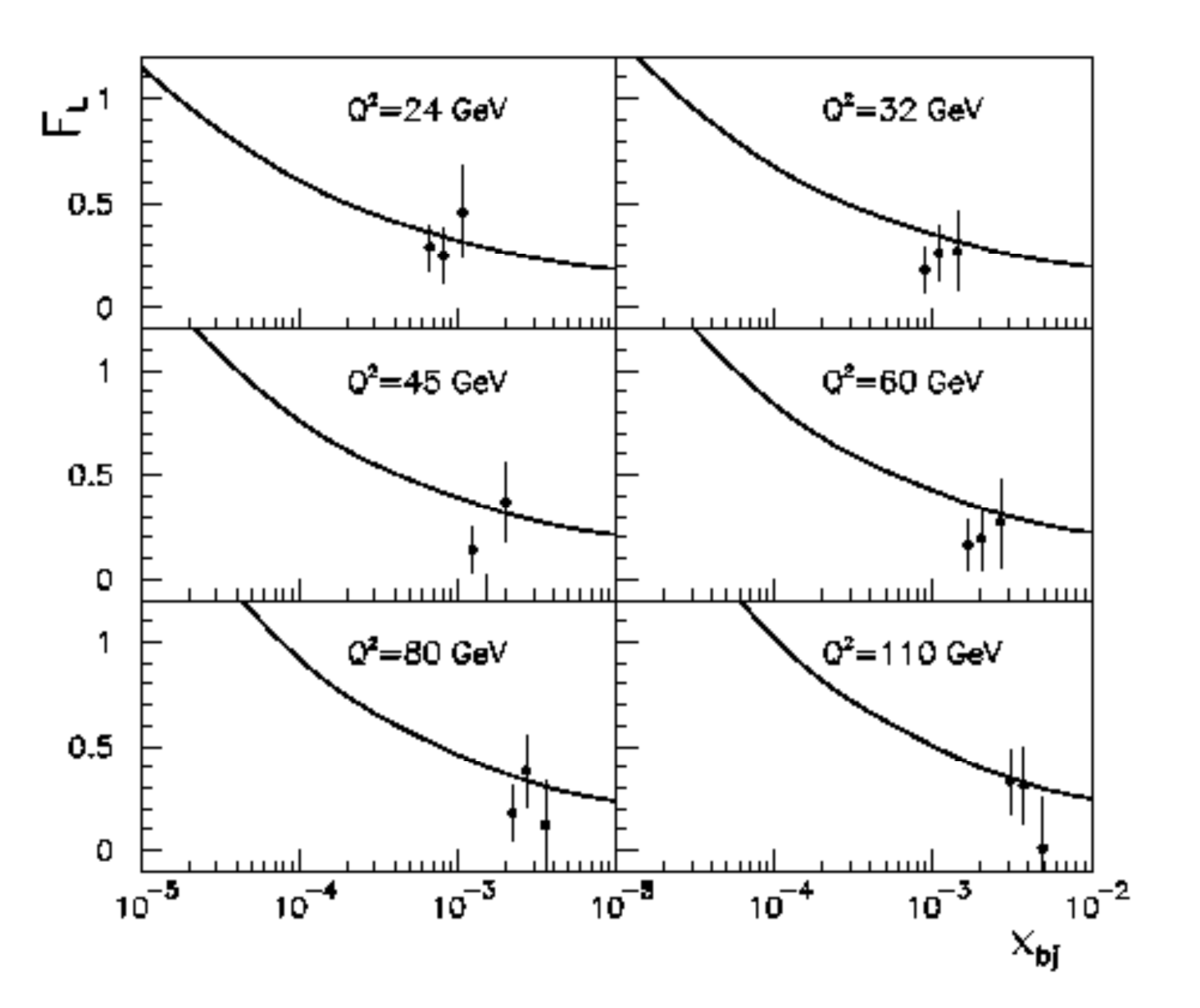}
\vspace*{-0.3cm}
\caption{The longitudinal proton structure function $F_L (x, Q^2)$
from the H1 collaboration \cite{h1data} and from the ZEUS collaboration
\cite{zeusdata} compared with the CDP prediction.}
\vspace*{-0.4cm}
\end{center}
\end{figure}

In Fig.\,19, we show the experimental results \cite{h1data, zeusdata} for
the longitudinal structure function, $F_L (x,Q^2)$, in
comparison with the predictions from the CDP based on the specific ansatz
for the dipole cross section that was used in Figs.\,15 to 18.


\section{Conclusions}

The color-gauge-invariant interaction of a $q \bar q$ dipole state
with the nucleon 
in terms of the forward scattering amplitude proceeds
via two reaction channels. They correspond to the two diagrams shown 
in Fig.\,9.
For a given dipole size, if both reaction channels are open, the 
cancellation between them implies {\it color transparency}. In the limit
in which the reaction
channel corresponding to the second diagram in Fig.\,9 is closed, the dipole
cross section {\it saturates} to a cross section of standard hadronic size with
at most a weak energy dependence.

In the photoabsorption cross section, the above limits of {\it color 
transparency and saturation} are realized, respectively, by the 
two regions in the $(Q^2, W^2)$ plane of Fig.\,2, corresponding to 
$\eta (W^2,Q^2) \gg 1$
and $\eta (W^2, Q^2) \ll 1$. The $(Q^2, W^2)$ plane (under the restriction
of $Q^2/W^2 \lsim ~0.1$) is accordingly simple. There are only two distinct
regions in the $(Q^2,W^2)$ plane,
separated by the line $\eta (W^2,Q^2) = 1$.

The main features of the experimental data on DIS at low x have thus been
recognized to follow from the color-gauge-invariant $q \bar q$
dipole-interaction with the gluon field of the proton
without adopting a specific ansatz for the dipole cross section.
Any specific ansatz has to interpolate between the model-independent
restrictions on the photoabsorption cross section that hold for
$\eta (W^2, Q^2) \gg 1$ and for
$\eta (W^2,Q^2) \ll 1$, respectively.

\section*{Acknowledgement}

The author thanks Kuroda-san for a fruitful collaboration on the
color dipole picture. Thanks to Prabhdeep Kaur for providing plots of
experimental data. The author thanks Allen Caldwell and Reinhart
K\"ogerler for useful discussions.

Last not least the author thanks Professor Antonino Zichichi for the
invitation to the 50th Anniversary Celebrations of his famous School of Subnuclear
Physics with its lively scientific atmosphere in the magnificent
environment of the ``Ettore Majorana'' Foundation and Centre for Scientific
Culture at Erice, Sicily.


\end{document}